\documentclass[a4paper,11pt]{article}
\usepackage[usenames]{color}

\usepackage{amsfonts,amssymb}
\usepackage{theorem}
\usepackage{epsf}
\usepackage[dvips]{graphicx}

\def\G{\Gamma}

\begin{document}
%\rightline{paper\_YM\_7.tex}

\rightline{IFUM-893-FT}

%\vskip 2.0 truecm
\Large
\bf
\centerline{A Massive Yang-Mills  Theory based}
\centerline{on the Nonlinearly Realized Gauge Group}
\normalsize \rm

\large
\rm
\vskip 1.3 truecm
\centerline{D.~Bettinelli
\footnote{e-mail: {\tt daniele.bettinelli@mi.infn.it}}, 
R.~Ferrari\footnote{e-mail: {\tt ruggero.ferrari@mi.infn.it}}, 
A.~Quadri\footnote{e-mail: {\tt andrea.quadri@mi.infn.it}}}

\normalsize
\medskip
\begin{center}
Dip. di Fisica, Universit\`a degli Studi di Milano\\
and INFN, Sez. di Milano\\
via Celoria 16, I-20133 Milano, Italy
\end{center}

\vskip 0.7  truecm
\normalsize
\bf
\centerline{Abstract}
\rm
\begin{quotation}
We propose a subtraction scheme for a massive Yang-Mills
theory realized via a nonlinear representation of the
gauge group (here SU(2)). It is based on the subtraction
of the poles in $D-4$ of the amplitudes, in dimensional
regularization, after a suitable normalization has been
performed. Perturbation theory is in the number of loops
and the procedure is stable under iterative subtraction
of the poles. The unphysical Goldstone bosons, the Faddeev-Popov
ghosts and the unphysical mode of the gauge field
are expected to cancel out in the unitarity equation.
The spontaneous symmetry breaking parameter is not
a physical variable. We use the tools already tested
in the nonlinear sigma model: hierarchy in the number of
Goldstone boson legs and weak power-counting property
(finite number of independent divergent amplitudes at each order).
It is intriguing that the model is naturally based
on the symmetry $SU(2)_L$ local $\otimes$ $SU(2)_R$
global. By construction the physical amplitudes depend
on the mass and on the self-coupling constant of the gauge particle
and moreover on the scale parameter of the radiative corrections.
The Feynman rules are in the Landau gauge.
\end{quotation}

\newpage

%%%%%%%%%%%%%%%%%%%%%%%%%%%%%%%%%%%%%%%%%
\section{Introduction}
%%%%%%%%%%%%%%%%%%%%%%%%%%%%%%%%%%%%%%%%%
%%%%%%%%%%%%%%%%%%%%%%%%%%%%%%%%%%%%%%%%%
\label{sec:intr}
With this work  we outline a theoretical framework
for the explicit evaluation of the Feynman amplitudes
of a massive Yang-Mills theory in its perturbative
loop expansion. We propose a subtraction scheme for the divergences
at $D=4$
and a robust set of symmetry requirements for the 
vertex functional in order to guarantee: stability under
the subtraction procedure, physical unitarity and predictivity.
\par
Quantization of non-abelian gauge theories is a subject with
a long history in quantum field theory. The perturbative
treatment of non-abelian gauge models was boosted by the
observation that the Yang-Mills action \cite{Yang:1954ek}
can be gauge-fixed
in such a way to guarantee  physical unitarity 
together with renormalizability by 
power-counting  (in the absence of anomalies)
\cite{Faddeev:1967fc,'tHooft:1971fh}. 
The discovery of the nilpotent BRST symmetry
\cite{brst} then provided a powerful and elegant tool
to study algebraically the gauge theories and in particular
physical unitarity to all orders in the perturbative
expansion \cite{physunit}. The implementation of the
BRST symmetry
by the Slavnov-Taylor (ST) identity \cite{ST} has boosted
unexpected progresses in quantum field theory (see e.g. \cite{Piguet:1995er}
and references therein).

As it is well-known, 
within this framework 
a mass term for the non-abelian gauge field
can be accounted for by enlarging the physical spectrum.
In fact the mass generation through spontaneous symmetry breaking (SSB) \cite{ssb}
in the presence of a linearly realized gauge symmetry
requires  the introduction of (at least) one physical scalar field, known as
the Higgs field. Power-counting renormalizability is preserved under 
this extension~\cite{renSSB}.

The latter field-theoretic paradigm has led to the extremely successful 
Standard Model of
particle physics. Still the question of the origin of SSB remains to be 
elucidated from the theoretical point of view, and the experimental
evidence of the existence of a Higgs particle is still waited for.

This paper is devoted to the analysis of a different approach to the 
subtraction of the divergences of the  massive Yang-Mills  theory
which relies on the use of a nonlinearly realized gauge group 
 through the introduction of a flat connection.
This strategy has  been applied in 
\cite{Ferrari:2005ii}-\cite{Bettinelli:2007kc} 
to the four-dimensional $SU(2)$ nonlinear sigma model. 
There the 
flat connection was coupled to an external vector source 
transforming as a background gauge field under 
the local $SU(2)_L$ left symmetry
\footnote{The left symmetry acts on the $SU(2)$ element  from the left.
In the following a global $SU(2)_R$ symmetry will also be introduced, 
acting on the group element from the right.}
which implements
the $SU(2)_L$-invariance of the  Haar measure in the path-integral.

\par
The present approach can be compared with the infinite
mass limit of the Higgs model in the linear case.
This has been already done in the case of the sigma model
in Ref.~\cite{Bettinelli:2006ps}. The same conclusions about the absence
of a general criterion for an unambiguous removal
of the $\log M_H$-parts apply here~\cite{Dittmaier:1995cr}.

\par
In a previous work \cite{Bettinelli:2007kc} we found a very powerful technique 
for integrating the functional equation derived from the invariance of
the path integral Haar measure under local $SU(2)_L$ transformations.
Our strategy in building a massive Yang-Mills  theory is based
on the same technique. We use the gauge field $A_\mu$ and the
nonlinear sigma model
field $\Omega$ to construct a {\sl bleached} gauge field $a_\mu$
\begin{eqnarray}
a_\mu \equiv \Omega^\dagger A_\mu \Omega -i\Omega^\dagger
\partial_\mu\Omega
\label{int.1}
\end{eqnarray}
which is invariant under $SU(2)_L$ transformations. Notice that each
element of the $2\times 2$ matrix is invariant. This opens far too many
possibilities than expected for constructing a gauge theory. 
In order to restrict
to the classical form of the massive Yang-Mills theory, we introduce some more
constraints. 
In particular we ask for global $SU(2)_R$ invariance (invariance
under local right $SU(2)$ transformation would forbid a mass term).
By this requirement all ``right'' indices are saturated and
consequently the number of invariants is drastically reduced.
This will be not enough. Therefore we will impose other constraints,
suggested by our previous works on the nonlinear sigma model.
They are aiming to control the severe 
divergences due to the presence of the nonlinear realization
of the gauge transformations: weak power-counting and hierarchy.
The first requirement controls the number of independent divergent amplitudes,
while the second guarantees that the amplitudes involving
the unphysical Goldstone field (descendant amplitudes) are determined
by the amplitudes of the ancestor fields (gauge fields, Faddeev-Popov
fields, composite fields associated to nonlinear transformations,
etc, i.e. most of the field content present in a power-counting renormalizable
gauge theory).

With this set of constraints we get a field theoretical model
in the Landau gauge which describes classically 
a massive non-abelian gauge field interacting with  the Faddeev-Popov ghosts
and non polynomially with the unphysical Goldstone bosons.
We stress that the model is BRST-, local $SU(2)_L$- and global
$SU(2)_R$-invariant and moreover it satisfies the necessary conditions
for the validity of the weak power-counting theorem.
We prove that the resulting equations for the 1-PI generating
functional (ST identity, local functional equation, ghost equation
and Landau gauge equation) are valid  for the amplitudes
constructed in $D$ dimensions by using the Feynman rules for the
loop expansion of the model (without any subtraction).
Moreover we demonstrate that minimal subtraction for the limit $D=4$
yields a consistent theory in terms of the parameters of the tree-level
effective action plus a mass scale for the radiative corrections.
The consistency of the theory relies upon some essential facts:
i) the subtraction of the divergences is achieved by local 
counterterms; ii) the number of the independent counterterms is finite at every
order of the loop expansion (as a consequence of the hierarchy
property and of the validity of the weak power-counting theorem);
iii) the subtraction procedure does not modify the defining equations;
iv) the validity of the ST identity guarantees the fulfillment
of physical unitarity. 
The last point requires that the Goldstone
bosons are unphysical modes together with the Faddeev-Popov ghosts
and the massless mode present in the Landau gauge description
of the vector field. 

Moreover it turns out that all the external sources coupled to
composite operators, which are necessary in order to perform
the subtraction of the divergences, are not physical parameters. 
In particular $K_0$, the source coupled to the order parameter field $\phi_0$
responsible of the spontaneous breakdown of the gauge symmetry,
is not physical. Then the physical amplitudes do not depend on
$v\equiv\langle \phi_0\rangle$.

The proof of physical unitarity (cancellation of unphysical states)
has been given
in Ref.~\cite{Ferrari:2004pd} both in the diagrammatic and in 
the operatorial formalism, 
under quite general assumptions 
which are fulfilled by the subtraction scheme discussed in the present paper.
The question of possible violations of the
Froissart unitarity bounds \cite{Froissart:1961ux, Horejsi:1993hz} 
that may occur at fixed perturbative order  
and the related issue of resummation of the perturbative
series will not be dealt with here.

It is somewhat important to investigate on the symmetry properties
of the counterterms by cohomological methods. For this purpose 
we consider the ST equation and the local functional equation
at the one loop level (the linearized ST- and local
functional-equations). The aim is to provide a basis for the
counterterms in terms
of local invariant solutions of these equations. These solutions
are parametrized by representatives of the cohomology of the
linearized ST operator on the space spanned by the 
local solutions of the linearized functional equation (i.e. the
variables {\sl ``bleached''} by a procedure similar to the one 
used in eq. (\ref{int.1})).  
\par
We have structured the paper according to the logical sequence
by which the requirements are imposed on the field theoretical model.
In Section \ref{sec:nlg} we construct the bleached fields according
to the nonlinear realization of the gauge group. The presence
of unwanted invariants suggests to impose the symmetry under global
$SU(2)_R$ transformations. In Section \ref{sec:wpc1}
the requirement of weak power-counting is imposed. 
In Section \ref{sec:brs1} the ST identity is derived
and it is shown that it is  not sufficient to yield the hierarchy.
In Section \ref{sec:chiral} we exploit the invariance of the
path integral measure under local 
gauge transformations and
derive the functional equation which yields both the hierarchy
and the subtraction procedure for the $D=4$ divergences.
In Section \ref{sec:brs2} we consider the final setup of all the
equations (ST identity, 
local functional equation, ghost equation, Landau gauge equation).
In Section \ref{sec:pert} we prove that the unsubtracted vertex
functional satisfies all the defining equations in the loop expansion.
The structure of the equations suggests the subtraction procedure
for the limit $D=4$. The equations are shown to be stable
after the introduction of the counterterms.
In Section \ref{sec:wpc2} we show that the whole set of identities
(ST identity, local functional equation, ghost equation,  Landau gauge
equation)
guarantees the hierarchy and thereby that Goldstone boson amplitudes
(descendant) are fixed by the ancestor amplitudes.
Section \ref{sec:unique} contains the implementation of
the weak power-counting to the construction of the tree level
vertex functional $\Gamma^{(0)}$ (massive Yang-Mills theory). 
In Section \ref{sec:one} we
discuss the properties of the local solutions of the linearized
equations and we list a complete set of them  compatible with
the required dimensions in the one-loop approximation.
The conclusions are in Section \ref{sec:concl}.
Appendix \ref{app:A} gives the Feynman rules, Appendix \ref{app:B}
proves that ST identity is not enough in order to impose the hierarchy
among the ancestor and  the descendant amplitudes, Appendix \ref{app:C}
yields the proof of the weak power-counting formula, Appendix \ref{app:D}
lists the linearized ST transforms of the bleached variables and
Appendix \ref{app:modST} is devoted to the proof of the
v.e.v.-independence of the physical amplitudes by using an extended ST
identity.

\section{Nonlinearly Realized Gauge Symmetries}
\label{sec:nlg}

The introduction in the Yang-Mills theory of a flat connection  
gives rise to a peculiar set of invariant variables which can be conveniently
described by making use of the technique discussed in \cite{Bettinelli:2007kc},
that we will briefly summarize here. It turns out that there
are many more invariants than in the usual approach based on $SU(2)$
local invariance mediated only by a vector meson. By adding extra fields
and in particular a flat connection one gets more terms. The usual
field strength term is achieved not only by requiring an invariance under
a large group, noticeably a global $SU(2)_R$ beside the local $SU(2)_L$,
but also by imposing the weak power counting criterion. This last
requirement will be dealt with later on.
\par
We will consider a $SU(2)$ gauge group and denote by 
$A_\mu = A_{a\mu} \frac{\tau_a}{2}$
the gauge connection. $\tau_a$ are the Pauli matrices.

The field strength of the gauge field $A_\mu$ is defined by
\begin{eqnarray}
G_{\mu\nu}[A] = G_{a\mu\nu} \frac{\tau_a}{2} =
\partial_\mu A_\nu - \partial_\nu A_\mu -i[A_\mu,A_\nu] \, .
\label{flc}
\end{eqnarray}
The nonlinear sigma model field $\Omega$ is an element of the $SU(2)$ group, 
which is parameterized in terms of the
coordinate fields $\phi_a$ as follows:
\begin{eqnarray}
&& \Omega = \frac{1}{v} ( \phi_0 + i \tau_a \phi_a ) \, , ~~~ \Omega^\dagger \Omega = 1 \,  , ~~~ {\rm det} \, \Omega = 1 \, ,
\nonumber \\
&& \phi_0^2 + \phi_a^2 = v^2  \, 
\label{s2.2}
\end{eqnarray}
where $v$ is a parameter with dimension equal one. We shall find out
that $v$ is not a parameter of the model, because it can can be
removed by a rescaling of the fields $\vec\phi, \phi_0$.
The $SU(2)$ flat connection is 
\begin{eqnarray}
&& F_\mu = i \Omega \partial_\mu \Omega^\dagger 
= F_{a\mu} \frac{\tau_a}{2} \, , \nonumber \\
&& F_{a\mu} = \frac{2}{v^2} (\phi_0 \partial_\mu \phi_a -
\partial_\mu \phi_0 \phi_a + \epsilon_{abc} \partial_\mu \phi_b \phi_c ) \, .
\label{s2.1}
\end{eqnarray}
The field strength of $F_\mu$ vanishes since $F_\mu$ is a flat connection
\begin{eqnarray}
G_{\mu\nu}[F] = 0 \, .
\label{flc.1}
\end{eqnarray}
Under a local $SU(2)$ left transformation 
$U_L = \exp \Big ( i \alpha^L_a \frac{\tau_a}{2} \Big )$ one gets
\begin{eqnarray}
&& \Omega' =  U_L \Omega \, , \nonumber \\
&& F'_\mu = U_L F_\mu U_L^\dagger + i U_L \partial_\mu U_L^\dagger \, , \nonumber \\
&& A'_\mu = U_L A_\mu U_L^\dagger + i U_L \partial_\mu U_L^\dagger \, .
\label{s2.4}
\end{eqnarray}
The nonlinearity of the $SU(2)$ constraint in eq.(\ref{s2.2})
implies that the gauge symmetry is nonlinearly realized on the 
fields $\phi_a$, whose
infinitesimal transformations are
\begin{eqnarray}
&& \delta \phi_a = \frac{1}{2} \phi_0 \alpha^L_a + \frac{1}{2} \epsilon_{abc} \phi_b \alpha^L_c \, , ~~~~
\phi_0 = \sqrt{v^2  - \phi_a^2} \, , \nonumber \\
&& \delta \phi_0 = -\frac{1}{2} \alpha^L_a \phi_a \, .
\label{s2.4.1}
\end{eqnarray}
 
Under local $SU(2)_L$ symmetry the combination $A_\mu - F_\mu$ transforms in the adjoint representation of $SU(2)$.
Hence one can construct out of $A_\mu - F_\mu$ and $\Omega$ a 
$SU(2)_L$-bleached variable $a_\mu$ which is invariant under $SU(2)_L$ local
transformations:
\begin{eqnarray}
a_\mu & = & a_{a\mu} \frac{\tau_a}{2} = 
            \Omega^\dagger (A_\mu - F_\mu) \Omega \nonumber \\
      & = & \Omega^\dagger A_\mu \Omega - i \partial_\mu \Omega^\dagger \Omega \, .
\label{s2.5}
\end{eqnarray}

The $SU(2)_L$ local symmetry is trivialized by the variable $a_\mu$, since any combination
of $a_\mu$ and its derivatives is $SU(2)_L$-invariant. 

One can also consider local $SU(2)_R$ transformations on $\Omega$
\begin{eqnarray}
\Omega' = \Omega U_R^\dagger 
\label{s2.6}
\end{eqnarray}
leaving $A_\mu$ invariant.

Then one finds that $a_\mu$ transforms as a $SU(2)_R$ gauge connection:
\begin{eqnarray}
a'_\mu = U_R  a_\mu U_R^\dagger + i U_R \partial_\mu U_R^\dagger \, .
\label{s2.7}
\end{eqnarray}

%%%%%%%%%%%%%%%%%%%%%%%%%%%%%%%%%%%%%%%%
%%%%%%%%%%%%%%%%%%%%%%%%%%%%%%%%%%%%%%%%
\subsection{Global $SU(2)_R$}

In the presence of a flat connection
the interplay of left  and right symmetries 
with renormalizability properties  provides very restrictive
constraints on the classical action.

In order to discuss this point we start from the Yang-Mills
action in the presence of a St\"uckelberg mass term \cite{st,Ferrari:2004pd}
\begin{eqnarray}
S & = &\frac{\Lambda^{(D-4)}}{g^2} \int d^Dx \, 
\Big ( -\frac{1}{4} G_{a\mu\nu}[ a] G_a^{\mu\nu}[ a] 
                     +\frac{M^2}{2} a_{a\mu}^2 \Big ) \nonumber \\
  & = &\frac{\Lambda^{(D-4)}}{g^2} \int d^Dx \, \Big ( - \frac{1}{4} 
                              G_{a\mu\nu}[A] G^{\mu\nu}_a[A] + 
                    \frac{M^2}{2} (A_{a\mu} - F_{a\mu})^2 \Big ) \, .
\label{stck.1}
\end{eqnarray}
$\Lambda$ is a mass scale for continuation in $D$ dimensions.

Notice that the field strength squared of 
$a_\mu$ coincides with the one of the gauge field $A_{a\mu}$
(since $a_\mu$ is obtained from $A_\mu$ through an operatorial
gauge transformation generated by $\Omega$).

$S$ is invariant under local $SU(2)_L$ symmetry (since it 
only depends on $a_\mu$) and also global $SU(2)_R$
symmetry. 
It is not invariant under local $SU(2)_R$
symmetry, since the latter forbids the St\"uckelberg mass
term because of the transformation property given in eq. (\ref{s2.7}).

Global $SU(2)_R$ symmetry restricts to some extent the number of independent
invariants (all right indices are saturated). 
We find it very intriguing that the symmetry under
global $SU(2)_R$ transformations is necessary in order to
reproduce a massive Yang-Mills gauge theory. In fact when one uses
the present theory for the electroweak model the $SU(2)_R$ global
symmetry plays the r\^ole of custodial symmetry \cite{fpq}. We stress that
our approach provides a natural justification of this property.

The implementation of the symmetries $SU(2)_L$ local and  $SU(2)_R$
global is also of great interest. From eq. (\ref{s2.4.1}) one
sees clearly that $SU(2)_L$ global is spontaneously broken
since the vacuum expectation
value of $\phi_0$ is non zero. The same conclusion is valid
for $SU(2)_R$ global. Thus only the symmetry generated by the
vector currents (L+R) is unitarely implemented and guarantees
a global $SU(2)$ symmetry for the physical amplitudes, while the symmetry
generated by the the axial currents is spontaneously broken.
This is another striking difference from massive Yang-Mills
realized in the realm of power counting renormalizable theories
\cite{gg}, where the $SU(2)$ local symmetry is spontaneously broken
in its global sector.

%%%%%%%%%%%%%%%%%%%%%%%%%%%%%%%%%%%%%%%%%%
%%%%%%%%%%%%%%%%%%%%%%%%%%%%%%%%%%%%%%%%%%
%%%%%%%%%%%%%%%%%%%%%%%%%%%%%%%%%%%%%%%%%%
\section{Weak Power-Counting I}
\label{sec:wpc1}

One should notice that global $SU(2)_R$ symmetry
allows for additional independent invariants 
which are also local $SU(2)_L$-symmetric.
For instance we have the following independent Lagrangian terms
of dimension $\leq 4$
\begin{eqnarray}
&
\int d^4x \, \partial_\mu a_{a\nu} \partial^\mu a^\nu_a \, , ~~~~
\int d^4x \, (\partial a)^2 \, ,  ~~~~ \int d^4x \, a^2 \, ,
& 
\nonumber \\
&
\int d^4x \, \epsilon_{abc} \partial_\mu a_{a\nu} a^\mu_b a^\nu_c \, , ~~~~
%\nonumber \\
%&
\int d^4x \, (a^2)^2 \, , ~~~~ \int d^4x \, a_{a\mu} a_{b}^\mu a_{a\nu}
a^\nu_b \, .  & 
\label{stck.2}
\end{eqnarray}
Thus the action $S$ 
in eq.(\ref{stck.1}) for  $D=4$ is not the most general 
Lorentz-invariant  functional with couplings 
of dimension $\ge 0$
compatible with local
$SU(2)_L$- and global $SU(2)_R$-symmetry.

However, $S$ 
is uniquely fixed
by local $SU(2)_L$-symmetry, global $SU(2)_R$-symmetry and
the requirement of weak power-counting property. By this
we mean that the number of superficially divergent independent
amplitudes is finite at each order in the loop expansion. 
This property is required to be stable under the procedure of 
subtraction of the divergences. While the second part of the statement
requires some effort, after the subtraction procedure has been 
given (see Section \ref{sec:pert}), the first part can be easily 
established,
under the assumptions discussed in Section \ref{sec:unique}. 
The proof of this central result requires to extend
the main tool 
developed to deal with the divergences 
of the nonlinear sigma model, i.e. the hierarchy  of the Feynman
amplitudes,
to the case where the  gauge bosons
are dynamical (see Appendix \ref{app:C}). 
\par
The weak power-counting property limits in a substantial
way the number of independent coefficients associated to the monomials
in eq. (\ref{stck.2}). One  observes that
each monomial in eq. (\ref{stck.2}) is a power series in the
Goldstone field $\vec\phi$ and moreover it contains in some cases
derivatives. The number of derivatives in the Goldstone 
interaction vertices is critical when one
evaluates the superficial degree of divergence of a graph.
Appendix \ref{app:A} provides some relevant Feynman rules
and Appendix \ref{app:C} gives the the superficial degree 
of divergence of a graph with no external  Goldstone lines.
In Section \ref{sec:unique} we prove that
the number of divergent ancestor amplitudes turns out to be a finite
only if the monomials of eq.(\ref{stck.2}) enter in the
combination given by the 
invariant $(G_a^{\mu\nu})^2$
and the presence of $\vec\phi$ is confined in the St\"uckelberg mass term.

%%%%%%%%%%%%%%%%%%%%%%%%%%%%%%%%%%%%%%%
%%%%%%%%%%%%%%%%%%%%%%%%%%%%%%%%%%%%%%%
%%%%%%%%%%%%%%%%%%%%%%%%%%%%%%%%%%%%%%%
\section{Slavnov-Taylor Identity  I}
\label{sec:brs1}

In order to set up the perturbative framework we use the Landau gauge. 

The gauge-fixing is performed by BRST techniques.
%This ensures Physical
%Unitarity  along the lines of~\cite{Ferrari:2004pd}.
%
The BRST differential $s$ is obtained in the usual way by promoting
the gauge parameters $\alpha_a^L$ to the ghost fields $c_a$ and by introducing 
the antighosts $\bar c_a$ coupled in a BRST doublet to the Nakanishi-Lautrup fields $B_a$:
\begin{eqnarray}
&& s \phi_a = \frac{1}{2} \phi_0 c_a + \frac{1}{2} \epsilon_{abc} \phi_b c_c \, , ~~~~ 
s A_{a\mu} = (D_\mu [A]c)_a \, , \nonumber \\
&& s \bar c_a = B_a \, , ~~~~ s B_a = 0 \, .
\label{brst.1}
\end{eqnarray}
In the above equation $D_\mu[A]$ denotes the covariant derivative w.r.t. $A_{a\mu}$:
\begin{eqnarray}
(D_\mu[A])_{ac} = \delta_{ac} \partial_\mu + \epsilon_{abc} A_{b\mu} \, .
\label{brst.1.1}
\end{eqnarray}
The BRST transformation of $c_a$ then follows by nilpotency 
\begin{eqnarray}
s c_a = - \frac{1}{2} \epsilon_{abc} c_b c_c \, .
\label{brst.2}
\end{eqnarray}

The tree-level vertex functional  is
\begin{eqnarray}
\G^{(0)} & = & S +\frac{\Lambda^{(D-4)}}{g^2}~ s~ \int d^Dx \, 
( \bar c_a \partial A_a )
\nonumber \\&&
+ \int d^Dx \, (A_{a\mu}^*  s A_a^\mu 
+ \phi_a^* s \phi_a + c_a^* s c_a ) \nonumber \\         
& = & S + \frac{\Lambda^{(D-4)}}{g^2}\int d^Dx \, \Big ( B_a \partial A_a - \bar c_a \partial_\mu (D^\mu[A] c)_a \Big ) \nonumber \\
&   & ~~~  + \int d^Dx \, (A_{a\mu}^*  s A_a^\mu + \phi_a^* s \phi_a + c_a^* s c_a ) \, .
\label{brst.3}
\end{eqnarray}
In $\G^{(0)}$ we have also included the antifields $A_{a\mu}^*, \phi_a^*$
and $c_a^*$ coupled to the nonlinear BRST variations of the
quantized fields.

We can assign a conserved ghost number by requiring that $A_{a\mu},\phi_a$ and $B_a$
have ghost number zero, $c_a$ has ghost number one, $\bar c_a, A_{a\mu}^*, \phi_a^*$
have ghost number $-1$ and finally $c_a^*$ has ghost number $-2$.
With these assignments the vertex functional has zero ghost number.

The propagators derived from $\G^{(0)}$ are collected in Appendix~\ref{app:A}.
>From eq.(\ref{prop.2}) one sees that the propagator for $\phi_a$ goes to infinity
like $1/p^{2}$. Since in $S$ there are interaction vertices with four $\phi$'s and two derivatives
(coming from the square of the flat connection), already at one loop level
there is an infinite number of divergent amplitudes with arbitrary number of $\phi$-legs.
This phenomenon is also present in the nonlinear sigma model and has been widely discussed in
Refs.\cite{Ferrari:2005ii}-\cite{Bettinelli:2007kc} .

In the nonlinear sigma model the way out is to make use of the hierarchy principle
\cite{Ferrari:2005ii} for the vertex functional, i.e. to fix the 
$\phi$-amplitudes in terms of ancestor amplitudes involving only the insertion
of the flat connection and the nonlinear sigma model constraint. 
This is achieved by making use of the local functional equation expressing the
invariance of the path-integral Haar measure under local $SU(2)_L$ transformations.
The number of divergent ancestor amplitudes is in turn
finite  at each order in perturbation theory 
(weak power-counting theorem) \cite{Ferrari:2005va}.

In the St\"uckelberg model the situation is somehow different. 
The invariance under the BRST symmetry in  eqs.(\ref{brst.1}),(\ref{brst.2})
can be translated into the following Slavnov-Taylor (ST) identity
\begin{eqnarray}&&
%\!\!\!\!\!\!\!\!\!\!
\!\!\!\!\!\!\!\!\!\!{\cal S}(\G^{(0)}) 
\nonumber\\&&
\!\!\!\!\!\!\!\!\!\!= \int d^Dx \, \Big (
\frac{\delta \G^{(0)}}{\delta A^*_{a\mu}} \frac{\delta \G^{(0)}}{\delta A_a^\mu}
+
\frac{\delta \G^{(0)}}{\delta \phi_a^*} \frac{\delta \G^{(0)}}{\delta \phi_a}
+ 
\frac{\delta \G^{(0)}}{\delta c_a^*}\frac{\delta \G^{(0)}}{\delta c_a}
+ B_a \frac{\delta \G^{(0)}}{\delta \bar c_a} \Big ) = 0 \, .
\label{brst.4}
\end{eqnarray}
This holds provided that the following dependence on the antifields of
the tree-level vertex functional $\G^{(0)}$ is imposed:
\begin{eqnarray}
&& \frac{\delta \G^{(0)}}{\delta A_{a\mu}^*} = (D^\mu[A]c)_a \, , \nonumber \\
&& \frac{\delta \G^{(0)}}{\delta \phi_a^*} = \frac{1}{2} \phi_0 c_a
+ \frac{1}{2} \epsilon_{abc} \phi_b c_c \, , \nonumber \\
&& \frac{\delta \G^{(0)}}{\delta c_a^*} = - \frac{1}{2} \epsilon_{abc} c_b c_c \, .
\label{tl.cond}
\end{eqnarray}

The ST identity for the full quantum vertex functional
is
\begin{eqnarray}
{\cal S}(\G) = 0 \, .
\label{st.1}
\end{eqnarray}
In power counting renormalizable theories the ST identity
is the tool to control the symmetry properties of the 
counterterms and to prove physical unitarity.
In the present case it has  some limitations, in particular
it does not imply the hierarchy property. 
In Appendix~\ref{app:B} an explicit counterexample is
fully developed. Here we give a short and simple argument.
We want to show that at least one particular amplitude,
involving only one $\vec\phi$ field cannot be obtained
by using eq. (\ref{st.1}) through the hierarchy mechanism.
These one-$\vec\phi$ field amplitudes can originate
only from the relevant term of the linearized eq. (\ref{st.1})
\begin{eqnarray}
\int d^Dx 
\frac{\delta\G^{(0)}}{\delta \phi_a^*(x)}
\frac{\delta}{\delta \phi_a(x)}\G^{(n)}.
\label{st.1.1.0}
\end{eqnarray}
Let us consider a one-loop amplitude given by the integrated monomial
\begin{eqnarray}
A_1\equiv\int d^Dy  A_{a\mu}^* c_b\partial^\mu\phi_c\epsilon_{abc}.
\label{st.1.1}
\end{eqnarray}
The action of the linearized ST operator in eq. (\ref{st.1.1.0}) 
connects a linearly $\vec\phi$-dependent amplitudes (as the example
in eq.(\ref{st.1.1})) to terms with no $\vec\phi$ (hierarchy).
We get
\begin{eqnarray}&&
\int d^Dx 
\frac{\delta\G^{(0)}}{\delta \phi_a^*(x)}
\frac{\delta}{\delta \phi_a(x)}
\int d^Dy  A_{a\mu}^* c_b\partial^\mu\phi_c\epsilon_{abc}
\nonumber\\&&
=v~\int d^Dx A_{a\mu}^* c_b
\partial^\mu c_c\epsilon_{abc} \dots
\nonumber\\&&
=\frac{1}{2}v~\int d^Dx A_{a\mu}^*\partial^\mu\left( c_b
c_c\right)\epsilon_{abc}\dots,
\label{st.1.2}
\end{eqnarray}
where dots represents terms with higher powers of $\vec\phi$,
which are irrelevant since we have to put $\vec\phi=0$.
Similarly  the monomial
\begin{eqnarray}
A_2\equiv\int d^Dy \partial^\mu A_{a\mu}^* c_b\phi_c\epsilon_{abc}
\label{st.1.3}
\end{eqnarray}
yields
\begin{eqnarray}&&
\int d^Dx 
\frac{\delta\G^{(0)}}{\delta \phi_a^*(x)}
\frac{\delta}{\delta \phi_a(x)}
\int d^Dy \partial^\mu A_{a\mu}^* c_b\phi_c\epsilon_{abc}
\nonumber\\&&
=v~\int d^Dx \partial^\mu A_{a\mu}^* c_b
c_c\epsilon_{abc} \dots.
\label{st.1.4}
\end{eqnarray}
Thus there is at least one amplitude that cannot be obtained
from the hierarchy procedure since
\begin{eqnarray}
\int d^Dx 
\frac{\delta\G^{(0)}}{\delta \phi_a^*(x)}
\frac{\delta}{\delta \phi_a(x)}
\biggl(2A_1+A_2
\biggr)\biggr|_{\vec\phi=0} =0.
\label{st.1.5}
\end{eqnarray}
Thus the set of ancestor fields (elementary or composite) have to be enlarged
in order fix completely the descendant amplitudes, i.e. those
involving one or more $\vec\phi$ field.
This will be done by using the functional equation that follows
from the invariance of the path integral measure under local gauge 
transformations.

%%%%%%%%%%%%%%%%%%%%%%%%%%%%%%%%%%%%%%%%%%%%%%%%%%%%
\section{Local Gauge Transformations}
\label{sec:chiral}
%%%%%%%%%%%%%%%%%%%%%%%%%%%%%%%%%%%%%%%%%%%%%%%%%%%%
%%%%%%%%%%%%%%%%%%%%%%%%%%%%%%%%%%%%%%%%%%%%%%%%%%%%
In order to overcome the difficulties arising from the 
absence of a hierarchy in the ST identity,
we make use of the local $SU(2)_L$ invariance
 of the path integral measure. While the classical action 
in eq. (\ref{stck.1}) is invariant under local 
gauge  transformations 
eq.(\ref{s2.4}), the gauge fixing term in eq.(\ref{brst.3}) is not.
In fact if we extend the gauge transformations 
(\ref{s2.4}) to the ghost fields by 
\begin{eqnarray}
&&
\delta c_a = \epsilon_{abc} c_b \alpha^L_c \, , 
\nonumber \\&&
\delta \bar c_a = \epsilon_{abc} \bar c_b \alpha^L_c 
\label{brst.-1}
\end{eqnarray}
we get 
\begin{eqnarray}
\delta S_{ GF} = -\Lambda^{D-4} \int d^D x ~\partial^\mu 
\alpha^{L}_a(x)~(s D_\mu[A]\bar{c})_a\, ,  
\label{brst.0}
\end{eqnarray}
where use has been made of the fact that the BRST differential $s$ and the 
 generator of infinitesimal gauge 
transformation $\delta$ are commuting operators  
\begin{eqnarray}
[s,\delta] =0\, .
\label{brst0.1}
\end{eqnarray}
In order to implement the gauge transformations properties for the 1-PI 
 vertex functional, we have to introduce a new set of external sources 
 coupled to the relevant composite operators. Thus the tree level 1-PI vertex
 functional becomes
\begin{eqnarray}
\G^{(0)} & = & S + 
\frac{\Lambda^{D-4}}{g^2}~
s \int d^Dx \, \Big(\bar c_a \partial^\mu A_{a\mu} \Big)  
\nonumber \\& & 
+ \frac{\Lambda^{D-4}}{g^2} 
\int d^Dx \, \Big(V_{a}^\mu~s (D_\mu[A]\bar{c})_a + 
\Theta_{a}^\mu~(D_\mu[A]\bar{c})_a\Big)\nonumber \\
& & + \int d^Dx \, \Big ( A^*_{a\mu} sA^\mu_a + 
\phi_0^* s \phi_0 + 
\phi_a^* s \phi_a + c_a^* s c_a
+ K_0 \phi_0 \Big )  \, .
\label{brst.4.1}
\end{eqnarray}
This can be recasted in the following form
\begin{eqnarray}
\G^{(0)} & = & 
 S + 
\frac{\Lambda^{D-4}}{g^2}
\int d^D x \, \Big ( B_a (D^\mu[V](A_\mu - V_\mu))_a
- \bar c_a (D^\mu[V] D_\mu[A] c)_a \Big ) 
\nonumber \\& & 
+ \frac{\Lambda^{D-4}}{g^2} \int d^Dx \, \Theta_{a}^\mu~(D_\mu[A]\bar{c})_a 
\nonumber \\& & 
+ \int d^Dx \, \Big ( A^*_{a\mu} sA^\mu_a + 
\phi_0^* s \phi_0 + 
\phi_a^* s \phi_a + c_a^* s c_a
+ K_0 \phi_0 \Big ) \, .
\label{brst.4.2}
\end{eqnarray}
The gauge fixing part can be interpreted as the background gauge fixing 
\cite{BFM} in the presence of the background connection $V_{a\mu}$.
\par
The tree level vertex functional in eq.(\ref{brst.4.2}) fulfills a local 
 functional equation which has to be preserved by the quantization
 procedure (which includes the subtraction of the divergences)
\begin{eqnarray}
&&{\cal W}(\G) \equiv \int d^Dx \alpha_a^L(x)\Biggl(
-\partial_\mu \frac{\delta \G}{\delta V_{a \mu}} 
+ \epsilon_{abc} V_{c\mu} \frac{\delta \G}{\delta V_{b\mu}}
-\partial_\mu \frac{\delta \G}{\delta A_{a \mu}}
\nonumber \\&&  
+ \epsilon_{abc} A_{c\mu} \frac{\delta \G}{\delta A_{b\mu}}
+ \epsilon_{abc} B_c \frac{\delta \G}{\delta B_b}
+ \frac{1}{2} K_0\phi_a
+ \frac{1}{2} \frac{\delta \G}{\delta K_0} 
\frac{\delta \G}{\delta \phi_a} 
\nonumber \\
&&  
+  
\frac{1}{2} \epsilon_{abc} \phi_c \frac{\delta \G}{\delta \phi_b} 
      + \epsilon_{abc} \bar c_c \frac{\delta \G}{\delta \bar c_b}
      + \epsilon_{abc} c_c \frac{\delta \G}{\delta  c_b}
 \nonumber \\
&& + \epsilon_{abc} \Theta_{c\mu} \frac{\delta \G}{\delta \Theta_{b\mu} }
      + \epsilon_{abc} A^*_{c\mu} \frac{\delta \G}{\delta A^*_{b\mu}}
      + \epsilon_{abc} c^*_c \frac{\delta \G}{\delta  c^*_b} 
 + \frac{1}{2} \phi_0^* \frac{\delta \G}{\delta \phi^*_a}
\nonumber \\
&& +  
\frac{1}{2} \epsilon_{abc} \phi^*_c \frac{\delta \G}{\delta \phi^*_b} 
- \frac{1}{2} \phi_a^* \frac{\delta \G}{\delta \phi_0^*} 
\Biggr)
= 0 \, .
\label{bkgwi}
\end{eqnarray}
The  interlacing between the local functional equation (\ref{bkgwi}) 
generated by the  gauge
 transformations and the ST identity will be treated in full detail in 
 the next Section. 
\par
We remark that the above equation contains a bilinear term, which arises as a 
 consequence of the nonlinearity of the local gauge transformations. 
This term allows  to establish the hierarchy procedure as in the 
nonlinear sigma model \cite{Ferrari:2005ii} , 
 \cite{Ferrari:2005va}, \cite{Bettinelli:2007kc}. 
The hierarchy tool allows to get 
 all the amplitudes involving at least one $\phi$ field 
(descendant amplitudes) from those 
 with no $\phi$ fields (ancestor amplitudes). 
The boundary condition for this algorithm is provided
by
\begin{eqnarray}
\frac{\delta \G}{\delta K_0(x)}\biggr|_{\rm All~fields~and~sources~=0}=v
\label{bou}
\end{eqnarray}
(see eq. (\ref{s2.4.1})). Since $\phi_0$ is not invariant both under
left- and right-$SU(2)$ transformations, both are spontaneously
broken by the condition (\ref{bou}) and only the $SU(2)_V$
is unitarely implemented. The parameter $v$ that breaks spontaneously
the symmetry is not a physical quantity. This can be seen
at the tree level in eqs. (\ref{brst.4.1}) and (\ref{brst.4.2})
where $v$ can be removed by the change of variables
\begin{eqnarray}&&
\G^{(0)}[ \vec A_\mu,\vec  c,\vec{  \bar c}, v\vec \phi,\vec  B, 
\vec A^*_\mu,\vec  c\,\,^*,  v^{-1}\vec \phi\,^*, 
v^{-1} \phi^*_0,  v^{-1} K_0, v]
\nonumber\\&&
=
\G^{(0)}[ \vec A_\mu,\vec  c,\vec{  \bar c}, \vec \phi,\vec  B, 
\vec A^*_\mu,\vec  c\,\,^*,  \vec \phi^*, \phi^*_0, K_0, v]\biggl|_{v=1}.
\label{bou.1}
\end{eqnarray}
or directly in eqs. (\ref{bkgwi}) and (\ref{bou}) where also 
$v$ disappears after one uses in $\G$ the substitution given 
in eq. (\ref{bou.1}). 
The rescaling of the field $\vec \phi$ has
no effects on the physical amplitudes.  
Also the effect of the rescaling on  the external sources $\phi^*_0$ and $K_0$
is null for Physics. However this conclusion can
be drawn only  after the enlargement of the ST transformations
to the new variables (Section \ref{sec:brs2}) and the
discovery that $\phi^*_0$ and   $K_0$ are not  physical variables. 

\par
In the sequel we will explicitly use the hierarchy
 procedure in the one-loop approximation by integrating 
the linearized form of eq.(\ref{bkgwi}) 
 as in Ref.~\cite{Bettinelli:2007kc}. Eq.(\ref{bkgwi}) together 
with the ST identity will be 
 our tool for the symmetric subtraction of the divergences in 
the perturbative expansion at the 
 point $D = 4$ for dimensionally regularized amplitudes. 
For that purpose we put in evidence  the linearized  part
of the above equation for future use both in the recursive 
construction of the counterterms in the loop expansion 
and in the integration over the variable $\vec \phi$. 
\begin{eqnarray} &&
{\cal W}_{0 }(\G^{(n)})  =   \int d^Dx \alpha_a^L(x)
\Biggl(-\partial_\mu \frac{\delta }{\delta V_{a \mu}} 
+ \epsilon_{abc} V_{c\mu} \frac{\delta }{\delta V_{b\mu}}
\nonumber \\&&
-\partial_\mu \frac{\delta }{\delta A_{a \mu}} 
+ \epsilon_{abc} A_{c\mu} \frac{\delta }{\delta A_{b\mu}}
 + \epsilon_{abc} B_c \frac{\delta }{\delta B_b}
      + \epsilon_{abc} \bar c_c \frac{\delta }{\delta \bar c_b}
\nonumber \\&& 
      + \epsilon_{abc} c_c \frac{\delta }{\delta  c_b}
+ \Big(\frac{1}{2}\delta_{ab}\frac{\delta \G^{(0)}}{\delta K_0}
+ \frac{1}{2} \epsilon_{abc} \phi_c 
\Big)\frac{\delta }{\delta \phi_b} + \frac{1}{2} 
\frac{\delta \G^{(0)}}{\delta \phi_a} \frac{\delta }{\delta K_0}
\nonumber \\&& 
+ \epsilon_{abc} \Theta_{c\mu} \frac{\delta }{\delta \Theta_{b\mu} }
      + \epsilon_{abc} A^*_{c\mu} \frac{\delta }{\delta A^*_{b\mu}}
      + \epsilon_{abc} c^*_c \frac{\delta }{\delta  c^*_b} \nonumber \\
&& + \frac{1}{2} \phi_0^* \frac{\delta }{\delta \phi^*_a} +  
\frac{1}{2} \epsilon_{abc} \phi^*_c \frac{\delta }{\delta \phi^*_b} 
- \frac{1}{2} \phi_a^* \frac{\delta }{\delta \phi_0^*} 
 \Biggr)\G^{(n)} 
\nonumber\\  && 
= -\frac{1}{2} \sum_{j=1}^{n-1}
\int d^Dx \alpha_a^L(x) \frac{\delta \G^{(j)}}{\delta K_0} 
\frac{\delta \G^{(n-j)}}{\delta \phi_a} \, .
\label{bkgwi.1}
\end{eqnarray} 
The requirement of the invariance under ${\cal W}_{0 a}$ corresponds 
to the invariance under the local
 transformations
%%%%%%%%%%%
%
\begin{eqnarray}
&& {\cal W}_0 A_{a\mu} = (D_\mu[A] \alpha^L)_a \, , ~~~~
   {\cal W}_0 V_{a\mu} = (D_\mu[V] \alpha^L)_a \, , \nonumber \\
&& {\cal W}_0 \phi_a = \frac{1}{2} \phi_0 \alpha^L_a 
+\frac{1}{2} \epsilon_{abc} 
   \phi_b \alpha^L_c \, , \nonumber \\
&& {\cal W}_0 B_a = \epsilon_{abc} B_b \alpha^L_c \, , \nonumber \\
&& {\cal W}_0 \bar c_a = \epsilon_{abc} \bar c_b \alpha^L_c \, , ~~~
   {\cal W}_0 c_a = \epsilon_{abc} c_b \alpha^L_c \, , \nonumber \\
&& {\cal W}_0 \Theta_{a\mu} = \epsilon_{abc} \Theta_{b\mu} \alpha^L_c \, , 
\nonumber \\
&& {\cal W}_0 A^*_{a \mu} = \epsilon_{abc} A^*_{b \mu}\alpha^L_c\, , ~~~~
   {\cal W}_0 c^*_a = \epsilon_{abc} c^*_b \alpha^L_c \, , \nonumber \\ 
&& {\cal W}_0 \phi_0^* = -\frac{1}{2} \alpha^L_a \phi_a^* \, , ~~~~
   {\cal W}_0 \phi_a^* = \frac{1}{2} \alpha^L_a \phi_0^*
+\frac{1}{2} \epsilon_{abc} \phi_b^* \alpha^L_c \, ,
\nonumber \\
&& {\cal W}_0 K_0 = \frac{1}{2}  \frac{\delta \G^{(0)}}{\delta\phi_a}  \alpha^L_a\, ,
\label{brst.5}
\end{eqnarray}
where
\begin{eqnarray}
{\cal W}_0 \equiv \int d^Dx \alpha_a^L(x){\cal W}_{0a} (x).
\label{brst.6}
\end{eqnarray}

The action of ${\cal W}_0$ on the fields coincides
with the one of the generator of the local gauge transformations.
In addition ${\cal W}_0$ also acts on the external sources,
as displayed in the last four lines of eq.(\ref{brst.5}).

The technique discussed in \cite{Bettinelli:2007kc} can be used
in order to derive a set of bleached variables (in one-to-one
correspondence with the original ones 
appearing in eq.(\ref{brst.5})) which are invariant
under ${\cal W}_0$. 

We first notice that 
for any  $I = I_a \frac{\tau_a}{2}$, transforming in the
adjoint representation under the local gauge
transformations in eq.(\ref{s2.4})
\begin{eqnarray}
I' = U_L I U_L^\dagger \, , 
\label{adj.1}
\end{eqnarray}
its bleached counterpart ${\widetilde I} = {\widetilde I}_a \frac{\tau_a}{2}$ can be obtained by conjugation w.r.t. $\Omega$
\begin{eqnarray}
{\widetilde I}  = \Omega^\dagger I \Omega \, .
\label{brst.6.3}
\end{eqnarray}
In fact  ${\widetilde I}$ is invariant under local gauge transformations.
In components one finds
\begin{eqnarray}
{\widetilde I}_a = R_{ba} I_b \, , 
\label{brst.6.4}
\end{eqnarray}
where the matrix $R_{ba}$ is given by
\begin{eqnarray}
R_{ba} \equiv\frac{1}{2} Tr \left(\Omega^\dagger \tau_b \Omega\tau_a\right)
=\Big ( 1 - 2 \frac{\vec{\phi}^2}{v_D^2} \Big )  \delta_{ba} 
              + 2 \frac{\phi_a \phi_b}{v_D^2} 
              + 2 \epsilon_{acb} \frac{\phi_0 \phi_c}{v_D^2} \, .
\label{brst.5.1}
\end{eqnarray}
This procedure allows to construct the 
bleached variables
\begin{eqnarray}
\widetilde{B}_a, \widetilde {\bar c}_a, \widetilde{c}_a,
\widetilde{\Theta}_{a\mu}, \widetilde{A^*}_{a\mu},
\widetilde{c^*_a} \, .
\label{brst.5.3}
\end{eqnarray}
Moreover, since $F_{a\mu}$ transforms as a flat connection under local
gauge transformations,  
the combinations $A_\mu-F_\mu$ and $V_\mu-F_\mu$,
both transform in  the adjoint representation.
The corresponding bleached variables are
denoted by $a_{a\mu}$ and $v_{a\mu}$ and are given by
\begin{eqnarray}
a_{a\mu} = R_{ba} (A_{b\mu} - F_{b\mu}) \, , ~~~~
v_{a\mu} = R_{ba} (V_{b\mu} - F_{b\mu}) \, .
\label{brst.6.2}
\end{eqnarray}
Since $R_{ba}$ is invertible, the change of variables
leading to the bleached variables in eqs.(\ref{brst.5.3})
and (\ref{brst.6.2}) is invertible.

We remark that all the bleached variables in eqs.(\ref{brst.5.3}) and (\ref{brst.6.2})  reduce  for $\phi=0$ to their corresponding  ancestors. One could also consider
the ${\cal W}_0$-invariant combination
\begin{eqnarray}
R_{ba} (A_{b\mu}  - V_{b\mu})
\label{comm.1}
\end{eqnarray}
but this would spoil the correspondence at $\phi=0$
with a single ancestor variable.

According to eq.(\ref{brst.5}) the matrix
\begin{eqnarray}
\Omega^* = \phi_0^* + i \phi^*_a \tau_a
\label{brst.7.1}
\end{eqnarray}
transforms  as $\Omega$ under ${\cal W}_0$.
In particular the combination 
$$ \Omega^\dagger \Omega^* = 
    \phi_0 \phi_0^* + \phi_a \phi_a^* +
    i (   \phi_a^* \phi_0 -\phi_0^* \phi_a - \epsilon_{abc} \phi_b^* 
   \phi_c) \tau_a 
$$
is ${\cal W}_0$-invariant.
This suggests to introduce the bleached counterparts
of $\phi_0^*$ and $\phi_a^*$ as follows:
\begin{eqnarray}
&& 
\!\!\!\!\!\!\!\!\!\!\!\!\!\!\!\!\!\!\!
\widetilde{\phi_0^*} = \frac{1}{v_D} (\phi_0 \phi_0^* +
\phi_a \phi_a^*)  \, , ~~~~
      \widetilde{\phi_a^*} = \frac{1}{v_D} (\phi_0 \phi_a^* - 
                            \phi_a \phi_0^* - \epsilon_{abc} \phi_b^* \phi_c ) \, .
\label{brst.7.2}
\end{eqnarray}
The normalization factor has been chosen in such a way 
that at $\phi=0$ $\widetilde{\phi_0^*}$ and
$\widetilde{\phi_a^*}$ reduce to
$\phi_0^*$ and $\phi_a^*$ respectively.

Finally it can be proved by the same methods used in
\cite{Ferrari:2005va} that the combination
\begin{eqnarray}
\widetilde{K_0} = 
\frac{1}{v_D} \Big ( \frac{v_D^2 K_0}{\phi_0} -
 \phi_a \frac{\delta}{\delta \phi_a} \Big (
 \left  . \G^{(0)} \right |_{K_0=0}  \Big ) \Big )  
\label{brst.7.3}
\end{eqnarray}
is ${\cal W}_0$-invariant. Again the normalization condition 
is chosen in such a way that $\left . \widetilde{K_0} \right |_{\phi=0} = K_0$ holds.

The use of the bleached variables will greatly simplify the solution
of the local functional equation (\ref{bkgwi.1}), since in these
variables ${\cal W}_0$ takes the very simple form
\begin{eqnarray}
{\cal W}_0 = \int d^Dx \, 
\alpha_b^L \zeta_{ab} \frac{\delta}{\delta \phi_a} \, , 
\label{brst.8}
\end{eqnarray}
where the invertible matrix $\zeta_{ab}$ is given by
\begin{eqnarray}
\zeta_{ab} = \frac{1}{2} \phi_0 \delta_{ab}
                 + \frac{1}{2} \epsilon_{acb} \phi_c \, .
\label{brst.9}
\end{eqnarray}
%

%
%%%%%%%%%%%%%%%%%%%%%%%%%%%%%%%%%%%%%%%%%%%%%%%%%%%%%%%%%  
%
%%%%%%%%%%%%%%%%%%%%%%%%%%%%%%%%%%%%%%%%%%%%%%%%%%%%%%%%%  

\section{Slavnov Taylor II}
\label{sec:brs2}
%%%%%%%%%%%%%%%%%%%%%%%%%%%%%%%%%%%%%%%%%%%%%%%%%%%%%%%%%  
%
%%%%%%%%%%%%%%%%%%%%%%%%%%%%%%%%%%%%%%%%%%%%%%%%%%%%%%%%%  

According to the standard algebraic treatment
given in \cite{grassi}-\cite{Ferrari:2000yp} the background connection $V_{a\mu}$
is paired with the classical ghost $\Theta_{a\mu}$
into a ${\cal S}_0$ doublet \cite{Barnich:2000zw}, \cite{Quadri:2002nh}:
\begin{eqnarray}
{\cal S}_0 V_{a\mu} = \Theta_{a\mu} \, , ~~~~
{\cal S}_0 \Theta_{a\mu} = 0 \, .
\label{brst.11}
\end{eqnarray}
This technical device allows to guarantee that physical
observables are not modified by the introduction
of the background connection \cite{Becchi:1999ir},\cite{Ferrari:2000yp}.
$\phi_0^*$ and $-K_0$ pair as well into a  ${\cal S}_0$ doublet:
\begin{eqnarray}
{\cal S}_0  \phi_0^* = -K_0 \, , ~~~ {\cal S}_0 K_0 = 0 \, .
\label{brst.12}
\end{eqnarray}
Under the assignments in eqs.(\ref{brst.11}) and
(\ref{brst.12}) $\G^{(0)}$ in eq.(\ref{brst.4.1}) is also
ST invariant.
We remark that, since the source $K_0$ 
of the nonlinear constraint in eq.(\ref{s2.2}) is the 
component of a  ${\cal S}_0$-doublet,
it is an unphysical variable (unlike in the nonlinear sigma model).
As a consequence the physical amplitudes are not affected
by the rescaling performed in eq. (\ref{bou.1}) and therefore they
do not depend from $v$.

The ST identity in the presence of the new set of sources
is
\begin{eqnarray}
&& \!\!\!\!\!\!\!\!\!\!
{\cal S}(\G) = \int d^Dx \, \Big (
\frac{\delta \G}{\delta A^*_{a\mu}} \frac{\delta \G}{\delta A_a^\mu}
+
\frac{\delta \G}{\delta \phi_a^*} \frac{\delta \G}{\delta \phi_a}
+ 
\frac{\delta \G}{\delta c_a^*}\frac{\delta \G}{\delta c_a}
+ B_a \frac{\delta \G}{\delta \bar c_a} \nonumber \\
&& ~~~~~~~~~~~~~~~~ + \Theta_{a\mu} \frac{\delta \G}{\delta V_{a\mu}}
      - K_0 \frac{\delta \G}{\delta \phi_0^*} 
\Big ) = 0 \, .
\label{brst.13}
\end{eqnarray}
$\G$ also obeys the Landau gauge equation
\begin{eqnarray}
\frac{\delta \G}{\delta B_a} = \frac{\Lambda^{D-4}}{g^2} 
 D^\mu[V](A_\mu - V_\mu)_a 
\label{b.eq}
\end{eqnarray}
and the ghost equation
\begin{eqnarray}
\frac{\delta \G}{\delta \bar c_a} = \frac{\Lambda^{D-4}}{g^2} 
 \Big ( 
-D_\mu[V] \frac{\delta \G}{\delta A_{\mu}^*} 
+ D_\mu[A] \Theta^\mu
\Big )_a \, ,
\label{gh.eq}
\end{eqnarray}
which follows as a consequence of the linearity of the gauge-fixing condition.
In the background Landau gauge a further identity holds, the 
antighost equation \cite{Grassi:2004yq}. However we will not make use of it
in the present construction since it cannot be generalized to different
Lorentz-covariant gauges.

The equations (\ref{brst.13}), (\ref{b.eq}) and (\ref{gh.eq}) are
not independent. By taking the functional derivative of eq. (\ref{brst.13})
with respect to $B$ and by using eq. (\ref{b.eq}) one obtains
the ghost equation (\ref{gh.eq}).

In the perturbative loop expansion we need to 
recursively use eq.(\ref{brst.13}) in order to
extract the symmetric counterterms. This leads us to consider
the linearized version of the ST identity
\begin{eqnarray}
&&
{\cal S}_0 ( \G^{(n)} )
\nonumber\\&& \equiv 
\int d^Dx \, \Big (
\frac{\delta \G^{(0)}}{\delta A^*_{a\mu}} \frac{\delta \G^{(n)}}{\delta A_a^\mu}
+
\frac{\delta \G^{(0)}}{\delta A_a^\mu}
\frac{\delta \G^{(n)}}{\delta A^*_{a\mu}} 
+
\frac{\delta \G^{(0)}}{\delta \phi_a^*} \frac{\delta \G^{(n)}}{\delta \phi_a}
+
\frac{\delta \G^{(0)}}{\delta \phi_a}
\frac{\delta \G^{(n)}}{\delta \phi_a^*} 
\nonumber \\
&& %~~~~~~~~ 
+
\frac{\delta \G^{(0)}}{\delta c_a^*}\frac{\delta \G^{(n)}}{\delta c_a}
+ 
\frac{\delta \G^{(0)}}{\delta c_a} \frac{\delta \G^{(n)}}{\delta c_a^*}
%\nonumber \\
%&& ~~~~~~~~
+ B_a \frac{\delta \G^{(n)}}{\delta \bar c_a}  + \Theta_{a\mu} \frac{\delta \G^{(n)}}{\delta V_{a\mu}}
      - K_0 \frac{\delta \G^{(n)}}{\delta \phi_0^*} 
\Big ) \nonumber \\
&& \!\!\!\!\!\!\!= -
\int d^Dx \, \, \sum_{j=1}^{n-1} \Big ( 
\frac{\delta \G^{(j)}}{\delta A^*_{a\mu}} \frac{\delta \G^{(n-j)}}{\delta A_a^\mu}
+
\frac{\delta \G^{(j)}}{\delta \phi_a^*} \frac{\delta \G^{(n-j)}}{\delta \phi_a}
+
\frac{\delta \G^{(j)}}{\delta c_a} \frac{\delta \G^{(n-j)}}{\delta c_a^*}
\Big )  \, .
\label{st.lin}
\end{eqnarray}

${\cal S}_0$ is nilpotent.

The Landau gauge equation (\ref{b.eq}) yields in the loop expansion at
order $n \geq 1$
\begin{eqnarray}
\frac{\delta \G^{(n)}}{\delta B_a} = 0 \, ,
\label{b.l.1}
\end{eqnarray}
i.e. the dependence on $B_a$ is only at tree-level.
Moreover the ghost equation (\ref{gh.eq}) 
yields at order $n \geq 1$
\begin{eqnarray}
\frac{\delta \G^{(n)}}{\delta \bar c_a} = 
\Lambda^{D-4} \Big ( 
- \partial_\mu \frac{\delta \G^{(n)}}{\delta A_{a\mu}^*} - \epsilon_{abc} V_{b\mu} 
\frac{\delta \G^{(n)}}{\delta A_{c\mu}^*} 
\Big ) 
\, .
\label{gh.eq.1.l}
\end{eqnarray}
The above equation implies that $\G^{(n)}$ depends 
on $\bar c_a$ only through the combination
\begin{eqnarray}
\widehat A^*_{a\mu} = A^*_{a\mu} + \Lambda^{D-4} (D_\mu [V] \bar c)_a \, .
\label{gh.eq.2.l}
\end{eqnarray}
The use of $\widehat A^*_{a\mu}$ instead of $A^*_{a\mu}$
simplifies the relevant ${\cal S}_0$-transforms involving 
$A^*_{a\mu}$. In fact one finds
\begin{eqnarray}
&& 
\!\!\!\!\!\!\!\!\!\!\!\!\!\!
{\cal S}_0 \widehat A^*_{a\mu} = 
\frac{\delta S}{\delta A_{a\mu}} - %\Lambda^{D-4} 
\epsilon_{abc} c_b \widehat A^*_{c\mu} \, , \nonumber \\
&& 
\!\!\!\!\!\!\!\!\!\!\!\!\!\!
{\cal S}_0 c_a^* = (D^\mu [A] \widehat A^*_\mu)_a
+ \frac{1}{2} \phi_0^* \phi_a - \frac{1}{2} \phi_a^* \phi_0 
- \frac{1}{2} \epsilon_{abc} \phi_b^*  \phi_c 
+ \epsilon_{abc} c_b^* c_c \, .
\label{gh.eq.3.l}
\end{eqnarray}
%
%%%%%%%%%%%%%%%%%%%%%%%%%%%%%%%%%%%%%%%%%%%%%%%%%%%%%%
\subsection{Physical and Unphysical  Quantities}
%%%%%%%%%%%%%%%%%%%%%%%%%%%%%%%%%%%%%%%%%%%%%%%%%%%%%%
With the transformation properties under ${\cal S}_0$ given
in this Section the only field that describes physical
states is $\vec A_\mu$. The massless mode of $\vec A_\mu$
(in the Landau gauge), the Goldstone bosons and the FP ghosts
are unphysical and are expected to give zero contribution
in the physical unitarity equation \cite{Ferrari:2004pd}.
Moreover also the external sources 
$A_{a\mu}^*, c_a^*, \phi_a^*,\phi_0^*, K_0$ are unphysical.
We stress once again the surprising fact that the
external source associated to the order parameter field
$\phi_0$ is not a physical variable.

The dependence on $v$ of the 1-PI vertex
functional can be discussed  by means of cohomological
tools as shown in Appendix~\ref{app:modST}. 
This is achieved by introducing an extended ST identity
under which also $v$ transforms into an anticommuting constant ghost $\theta$.
This extended ST identity holds for the quantum effective
action whose classical approximation $\G_{ext}^{(0)}$ 
 involves an additional $\theta$-dependent part.
 $\G_{ext}^{(0)}$ reduces for $\theta=0$
to $\G^{(0)}$ in eq.(\ref{brst.4.2}).
The advantage of this procedure is that it allows to discuss
the dependence on $v$ of the connected Green functions
by  algebraic methods which are close to those developed
in gauge theories in order to discuss the dependence
of the connected generating functional on the 
gauge parameter \cite{Piguet:1984js}.
One finds that the connected Green functions of BRST-invariant
local operators are independent of $v$. 
This is a rather remarkable result, since it shows that 
in the present approach $v$ is an unphysical mass scale.
Moreover one
can derive an equation allowing to control the dependence
of the Green functions involving $K_0$ in terms
of those involving the antifield $\phi_0^*$. This reflects
the fact that $\phi_0^*$ and $-K_0$ form 
a ${\cal S}_0$-doublet (see eq.(\ref{brst.12})).
In this connection we remark that  
the issue of whether the composite operator $\phi_0$,
coupled to the external source $K_0$, is physical or not
is a somewhat peculiar problem. By standard cohomological
arguments \cite{Barnich:2000zw} it can be proved that $\phi_0^*$ and 
$K_0$ do not contribute to the cohomology of the linearized
ST operator ${\cal S}_0$ (because they form a  ${\cal S}_0$-doublet
\cite{Barnich:2000zw}, \cite{Quadri:2002nh}).
Since in the perturbation expansion of gauge theories
physical observables can be identified with the 
cohomology classes of ${\cal S}_0$, 
we conclude that $K_0$ is unphysical.
%%%%%%%%%%%%%%%%%%%%%%%%%%%%%%%%%%%%%%%%%%%%%%%%%%%%%%
\section{Perturbative Solution in $D$ Dimensions}
%%%%%%%%%%%%%%%%%%%%%%%%%%%%%%%%%%%%%%%%%%%%%%%%%%%%%%
%%%%%%%%%%%%%%%%%%%%%%%%%%%%%%%%%%%%%%%%%%%%%%%%%%%%%%
\label{sec:pert}
It is of paramount importance to establish whether
eqs.  (\ref{bkgwi}), (\ref{brst.13}) and (\ref{b.eq})
are compatible. For our purpose it would be very
satisfactory  to prove that the perturbative expansion in the number of
loops of the generating functional of the 1-PI functions
yields a solution of both equations.
This
is indeed the case and  the proof
of this result is very close to the one already
given for the nonlinear sigma model in Ref. \cite{Bettinelli:2007zn}. 
Thus we will not repeat it here. 
The Feynman rules are taken from the classical action
in eq. (\ref{brst.4.2}) in $D$ dimensions. It should be noticed a technical
point regarding the presence of massive tadpoles. In 
dimensional regularization they are non zero, unlike in the massless
case. Therefore one should keep track of them.
\par
Let us state here only the final formulas. In the present Section
we perform a rescaling by a factor
\begin{eqnarray}
\Lambda_D \equiv \frac{\Lambda^{(D-4)}}{g^2}
\label{pert.-1}
\end{eqnarray}
of the anti-fields $A_{a\mu}^*, c_a^*, \phi_a^*$ and of the external
sources $\phi_0^*, K_0$
\begin{eqnarray}
\biggl(A_{a\mu}^*, c_a^*, \phi_a^*,\phi_0^*, K_0\biggr)
\to \Lambda_D\biggl(A_{a\mu}^*, c_a^*, \phi_a^*,\phi_0^*, K_0\biggr)
\label{pert.0}
\end{eqnarray}
so that the unperturbed effective action ((\ref{brst.4.1}) or (\ref{brst.4.2}))
becomes
\begin{eqnarray}
\G^{(0)} & = & S + \Lambda_D
~
s \int d^Dx \, \Big(\bar c_a \partial^\mu A_{a\mu} \Big)  
\nonumber \\& & 
+ \Lambda_D
\int d^Dx \, \Big(V_{a}^\mu~s (D_\mu[A]\bar{c})_a + 
\Theta_{a}^\mu~(D_\mu[A]\bar{c})_a\Big)\nonumber \\
& & + \Lambda_D\int d^Dx \, \Big ( A^*_{a\mu} sA^\mu_a + 
\phi_0^* s \phi_0 + 
\phi_a^* s \phi_a + c_a^* s c_a
+ K_0 \phi_0 \Big )  \, .
\label{pert.0.0}
\end{eqnarray}
This rescaling is introduced in order to give $D-$independent
canonical dimensions to all the ancestor fields and sources.
It introduces however some $\Lambda_D$-dependent factors
both in the local gauge functional equation and in the
ST identity. We shall account for this change, since it is
important for the subtraction procedure. Moreover we denote by 
\begin{eqnarray}
\widehat\G\equiv \Gamma^{(0)}
+\sum_{j\geq 1}\widehat\Gamma^{(j)}.
\label{mod.12}
\end{eqnarray}
the whole set of Feynman rules, including the
counterterms. 
The local functional $\widehat \G^{(j)}$ collects
all the counterterms of order $\hbar^j$.
%
%%%%%%%%%%%%%%%%%%%%%%%%%%%%%%%%%%%%%%%%%
\subsection{Local Gauge Equation}
In generic $D$ dimensions after the rescaling of
  eq. (\ref{pert.0}) the
functional $\textit{Z}$, generating the Feynman amplitudes, obeys
the equation associated to the local gauge transformations
\begin{eqnarray}&&
\Biggl(-\partial_\mu \frac{\delta }{\delta V_{a \mu}} 
+ \epsilon_{abc} V_{c\mu} \frac{\delta }{\delta V_{b\mu}}
+\partial_\mu L_a^ \mu 
- \epsilon_{abc}  L_{b \mu} \frac{\delta }{\delta L_{c\mu}} 
- \epsilon_{abc}J^B_b\frac{\delta }{\delta J^B_c}  
\nonumber \\
&&
      + \epsilon_{abc}\eta_b   \frac{\delta }{\delta \eta_c}
      + \epsilon_{abc} \bar\eta_b\frac{\delta }{\delta  \bar\eta_c} 
+\frac{\Lambda_D}{2}   K_0  \frac{\delta }{\delta K_a} 
 -\frac{1}{2\Lambda_D} K_a \frac{\delta }{\delta K_0}
- \frac{1}{2} \epsilon_{abc}K_b \frac{\delta }{\delta K_c}
\nonumber \\
&& + \epsilon_{abc} \Theta_{c\mu} \frac{\delta }{\delta \Theta_{b\mu} }
      + \epsilon_{abc} A^*_{c\mu} \frac{\delta }{\delta A^*_{b\mu}}
      + \epsilon_{abc} c^*_c \frac{\delta }{\delta  c^*_b} 
\nonumber \\
&& + \frac{1}{2} \phi_0^* \frac{\delta }{\delta \phi^*_a} +  
\frac{1}{2} \epsilon_{abc} \phi^*_c \frac{\delta }{\delta \phi^*_b} 
- \frac{1}{2} \phi_a^* \frac{\delta }{\delta \phi_0^*} 
 \Biggr)\textit{Z}
\nonumber \\&&  
 =i \Biggl(-\partial_\mu \frac{\delta \widehat\G}{\delta V_{a \mu}} 
+ \epsilon_{abc} V_{c\mu} \frac{\delta \widehat\G}{\delta V_{b\mu}}
-\partial_\mu \frac{\delta \widehat\G}{\delta A_{a \mu}} 
+ \epsilon_{abc} A_{c\mu} \frac{\delta \widehat\G}{\delta A_{b\mu}}
 + \epsilon_{abc} B_c \frac{\delta\widehat\G }{\delta B_b}
\nonumber \\
&&
      + \epsilon_{abc} \bar c_c \frac{\delta\widehat\G }{\delta \bar c_b}
      + \epsilon_{abc} c_c \frac{\delta\widehat\G }{\delta  c_b} 
+\frac{\Lambda_D}{2}  K_0 \phi_a 
+\frac{1}{2\Lambda_D}\frac{\delta \widehat\G}{\delta K_0} 
\frac{\delta \widehat\G}{\delta \phi_a}
+\frac{1}{2} \epsilon_{abc}\phi_c \frac{\delta \widehat\G}{\delta \phi_b} 
\nonumber \\
&& + \epsilon_{abc} \Theta_{c\mu} \frac{\delta \widehat\G}{\delta \Theta_{b\mu} }
      + \epsilon_{abc} A^*_{c\mu} \frac{\delta\widehat\G }{\delta A^*_{b\mu}}
      + \epsilon_{abc} c^*_c \frac{\delta\widehat\G }{\delta  c^*_b} \nonumber \\
&& + \frac{1}{2} \phi_0^* \frac{\delta \widehat\G}{\delta \phi^*_a} +  
\frac{1}{2} \epsilon_{abc} \phi^*_c \frac{\delta\widehat\G }{\delta \phi^*_b} 
- \frac{1}{2} \phi_a^* \frac{\delta \widehat\G}{\delta \phi_0^*} 
\Biggr)\cdot \textit{Z}\, ,
\label{mod.11}
\end{eqnarray}
where the dot indicates the insertion of the local operators and the field
sources are given by

\begin{eqnarray}&&
L_{a\mu} = -\frac{\delta\Gamma}{\delta A^\mu_a}\qquad
K_a = -\frac{\delta\Gamma}{\delta \phi_a}\qquad
J^B_a = -\frac{\delta\Gamma}{\delta B_a}\qquad
\nonumber\\&&
\eta_a = -\frac{\delta\Gamma}{\delta \bar c_a}\qquad
\bar\eta_a = \frac{\delta\Gamma}{\delta  c_a}.
\label{pert.1}
\end{eqnarray}
If no counterterms are present then $\widehat\Gamma=\Gamma^{(0)}$,
then eq. (\ref{mod.11}) proves that the unsubtracted amplitudes
in $D$ dimensions satisfy the functional equation associated to
the local gauge transformations. In fact $\Gamma^{(0)}$ is
by construction a solution of eq. (\ref{bkgwi}) and therefore the
R.H.S. of eq. (\ref{mod.11}) is zero. On the other side, if counteterms
are introduced, they must obey the identity
\begin{eqnarray}&&
-\partial_\mu \frac{\delta \widehat\G}{\delta V_{a \mu}} 
+ \epsilon_{abc} V_{c\mu} \frac{\delta \widehat\G}{\delta V_{b\mu}}
-\partial_\mu \frac{\delta \widehat\G}{\delta A_{a \mu}} 
+ \epsilon_{abc} A_{c\mu} \frac{\delta \widehat\G}{\delta A_{b\mu}}
 + \epsilon_{abc} B_c \frac{\delta\widehat\G }{\delta B_b}
\nonumber \\
&&
      + \epsilon_{abc} \bar c_c \frac{\delta\widehat\G }{\delta \bar c_b}
      + \epsilon_{abc} c_c \frac{\delta\widehat\G }{\delta  c_b} 
+\frac{\Lambda_D}{2}  K_0 \phi_a 
+\frac{1}{2\Lambda_D}\frac{\delta \widehat\G}{\delta K_0} 
\frac{\delta \widehat\G}{\delta \phi_a}
+\frac{1}{2} \epsilon_{abc}\phi_c \frac{\delta \widehat\G}{\delta \phi_b} 
\nonumber \\
&& + \epsilon_{abc} \Theta_{c\mu} \frac{\delta \widehat\G}{\delta \Theta_{b\mu} }
      + \epsilon_{abc} A^*_{c\mu} \frac{\delta\widehat\G }{\delta A^*_{b\mu}}
      + \epsilon_{abc} c^*_c \frac{\delta\widehat\G }{\delta  c^*_b} \nonumber \\
&& + \frac{1}{2} \phi_0^* \frac{\delta \widehat\G}{\delta \phi^*_a} +  
\frac{1}{2} \epsilon_{abc} \phi^*_c \frac{\delta\widehat\G }{\delta \phi^*_b} 
- \frac{1}{2} \phi_a^* \frac{\delta \widehat\G}{\delta \phi_0^*}=0.
\label{pert.2}
\end{eqnarray}
% 
%
%%%%%%%%%%%%%%%%%%%%%%%%%%%%%%%%%%%%%%%%%
\subsection{The Subtraction Procedure}

Eq. (\ref{pert.2}) is the tool used in order to construct the counterterms
necessary for the limit $D=4$. Assume that the subtraction procedure
has been performed up to order $n-1$. Only the pole parts are subtracted
by adopting the counterterms structure
\begin{eqnarray}
\widehat\G= \Gamma^{(0)}+ \Lambda_D
\sum_{j\ge 1}\int d^Dx {\cal M}^{(j)},
\label{pert.3}
\end{eqnarray}
where the local polynomials ${\cal M}^{(j)}$ in the fields
and sources have no $D$ dependence apart from the poles in $D-4$.
At order $n$  eq. (\ref{bkgwi.1}) is then violated since
the $n-$th countertems are not present as they should according
to eq. (\ref{pert.2}). The violation is explicitly given by
\begin{eqnarray}
{\cal W}_{0 a}(\G^{(n)}) 
+\frac{1}{2\Lambda_D} \sum_{j=1}^{n-1} \frac{\delta \G^{(j)}}{\delta K_0} 
\frac{\delta \G^{(n-j)}}{\delta \phi_a} 
% 
%\nonumber\\&&
 = 
 \frac{1}{2\Lambda_D} \sum_{j=1}^{n-1}
\frac{\delta \widehat\G^{(j)}}{\delta K_0} 
\frac{\delta \widehat\G^{(n-j)}}{\delta \phi_a}.
\label{pert.4}
\end{eqnarray} 

According to eq. (\ref{pert.3}) the pole part ${\cal M}^{(n)}$ has to be
extracted from the normalized amplitude
\begin{eqnarray}
\Lambda_D^{-1} \Gamma^{(n)}.
\label{pert.5.2}
\end{eqnarray}
By this normalization condition the R.H.S. in eq. (\ref{pert.4})
is $D$-independent apart from the poles in $D-4$ by construction (as stated
in eq. (\ref{pert.3})). Then minimal subtraction on the normalized
amplitude (\ref{pert.5.2}) removes  the breaking terms
in the R.H.S. In the L.H.S. of eq. (\ref{pert.4}) ${\cal W}_{0 a}$ at $n >
0$ contains no $\Lambda_D$ factor and therefore the procedure of
subtraction (normalization according to eq. (\ref{pert.5.2}) and pure
pole subtraction) does not modify the equation.

Further details about this subtraction procedure are in Appendix D
of Ref. \cite{Bettinelli:2007zn}. 

Once again we stress
our point of view that, by using the freedom to introduce free
parameters describing the general solution $\Delta\widehat\Gamma^{(n)}$
of the homogeneous equation
\begin{eqnarray}
{\cal W}_0(\Delta\widehat\Gamma^{(n)})=0,
\label{pert.6}
\end{eqnarray}
one would destroy the predictivity of the theory, since the theory 
is not power counting renormalizable 
and therefore the new parameters
appearing in the quantum corrections  cannot be reabsorbed
by a redefinition of the constants already present in the classical
vertex functional $\G^{(0)}$. Our subtraction prescription
is based on a finite number of parameters. Therefore it is  predictive
and it can be experimentally tested.

\subsection{Comments on the Subtraction Procedure}
Let us look closer into this subtraction procedure, by considering
the dependence from $\Lambda_D$ of a generic amplitude. 
The removal of the divergences requires the insertion of
counterterms. 
Therefore it is important to distinguish the order in the $\hbar$ expansion
from the loop number. A counterterm ${\cal M}^{(k)}$ in eq. (\ref{pert.3})
is of order $\hbar^k$. The vertex functional can be graded according
to the $\hbar$ power of the counterterms included in the amplitudes
(in $D$ dimensions):
\begin{eqnarray}
\Gamma^{(n)} = \sum_{k=0}^n \Gamma^{(n,k)}.
\label{pert.6.1}
\end{eqnarray}
$\Gamma^{(n,k)}$ has important properties that are discussed in
Ref. \cite{Bettinelli:2007zn}.
\par
With the Feynman rules given by the $\G^{(0)}$ in eq.
(\ref{pert.0.0}) the propagators of the dynamical fields
carry a factor $\Lambda_D^{-1}$ while every vertex
has a factor $\Lambda_D$ (including the counterterms). 
Since  the number $n_L$ of topological loops for a 1PI amplitude is given by
\begin{eqnarray}
n_L = I -V + 1,
\label{pert.5.0}
\end{eqnarray}
where $V$ is the number of vertices 
(including the counterterms). Then the factor is
\begin{eqnarray}
\Lambda_D^{(1-n_L)}.
\label{pert.5.1p}
\end{eqnarray}
% 
%\par\noindent
Therefore  $\Gamma^{(n,k)}$ carries an overall factor
given by a power of $\Lambda_D$, where the exponent is not 
given by the order in the $\hbar$ expansion, but by the number
of topological loops
\begin{eqnarray}
\Gamma^{(n)} = \sum_{k=0}^n\Lambda_D^{(1-n_L)}\Gamma^{(n,k)}
\biggl|_{\Lambda_D=1}
=\Lambda_D^{(1-n)}\sum_{k=0}^n\Lambda_D^{k}\Gamma^{(n,k)}
\biggl|_{\Lambda_D=1}
\label{pert.5.3p}
\end{eqnarray}
where $\Lambda_D$ can be set to one, by considering it as an
independent variable together with $\Lambda$ and $D$:
\begin{eqnarray}
g^2 = \frac{\Lambda^{(D-4)}}{\Lambda_D}.
\label{pert.5.2pp}
\end{eqnarray}
The power of $\hbar^{n}$ of $\Gamma^{(n,k)}$ is given by
\begin{eqnarray}
n= I + \sum_{j\ge 0} V^{(j)}(j-1) +1= I-V + 1 + k
= n_L + k,
\label{pert.5.2p}
\end{eqnarray}
where $ V^{(j)}$ counts the number of vertices of order  $\hbar^{j}$
and $k$ is the total $\hbar$-power of the counterterms.
In particular the tree-vertices are of order $\hbar^{0}$.
Also the coupling constant  enters in the amplitudes  in a 
power-like form  ($g^{2(n-1)}$), since the subtraction procedure does not
alter the dependence on $g$.

The complex dependence of the vertex functional from $\Lambda_D$ makes
the subtraction procedure non trivial. In the iterative procedure
of subtraction, where the counterterms
have been consistently used up to order $n-1$, 
the 1PI amplitude $\Gamma^{(n)}_U$ (where the subscript $U$ reminds
that the last subtraction at order $n$ has yet to be performed)
has a Laurent expansion
\begin{eqnarray}
\Gamma^{(n)}_U=\sum_{j=-M}^\infty a_j (D-4)^j
\label{pert.5.4}
\end{eqnarray}
Then the proposed 
finite part is given by the $(D-4)^0$ coefficient in the Laurent expansion
of $\Lambda^{(4-D)}\Gamma^{(n)}_U$. I.e.
\begin{eqnarray} 
\sum_{j=0}^{M}\frac{1}{j!}(-\ln(\Lambda))^j~a_{-j}.
\label{pert.5.5}
\end{eqnarray}
While the counterterms are given by
\begin{eqnarray}&& \int d^Dx 
{\cal M}^{(n)}(x)
= -g^2\sum_{i=0}^\infty \frac{1}{i!}(-\ln(\Lambda))^i (D-4)^i
\sum_{j=0}^{M}a_{-j} (D-4)^{-j}\biggr|_{\rm Pole~Part}
\nonumber\\&&
=  -g^2\sum_{l=1}^M\frac{1}{(D-4)^{l}}
\biggl( \sum_{j=l}^{M}\frac{1}{(j-l)!}(-\ln(\Lambda))^{(j-l)} 
a_{-j} \biggr).
\label{pert.5.6p}
\end{eqnarray}
One can easily verify that the finite part for $D=4$ of
\begin{eqnarray}&&
\Gamma^{(n)}_U + \Lambda_D\int d^Dx {\cal M}^{(n)}(x)\biggr|_{D=4}
=\Lambda_D\biggl(\frac{1}{\Lambda_D}\Gamma^{(n)}_U 
+\int d^Dx {\cal M}^{(n)}(x)\biggr)\biggr|_{D=4}
\nonumber\\&&
=\frac{1}{g^2}
\biggl(\frac{1}{\Lambda_D}\Gamma^{(n)}_U 
+\int d^Dx {\cal M}^{(n)}(x)\biggr)\biggr|_{D=4}
\label{pert.5.7}
\end{eqnarray}
is indeed the expression given in eq. (\ref{pert.5.5}).

%
%%%%%%%%%%%%%%%%%%%%%%%%%%%%%%%%%%%%%%%%%
\subsection{Slavnov-Taylor Equation}
Now we examine the same items for the ST identity (\ref{brst.13}). 
The ghost equation (\ref{gh.eq}),
being linear in $\Gamma$, poses no problems.
\par 
As for the functional equation (\ref{mod.11}) associated to
the local gauge transformations, we state the relation
between ST identity and the equation for the counterterms
\begin{eqnarray}
&& 
 \int d^Dx \, \Biggr (-\frac{L_{a\mu}}{\Lambda_D}
\frac{\delta}{\delta A^*_{a\mu}} 
-\frac{K_a}{\Lambda_D}
\frac{\delta }{\delta \phi_a^*} 
+ 
\frac{\bar\eta_a}{\Lambda_D}\frac{\delta }{\delta c_a^*}
-\eta_a  \frac{\delta }{\delta J^B_a} 
\nonumber \\&&  
+ \Theta_{a\mu} \frac{\delta }{\delta V_{a\mu}}
      -  K_0 \frac{\delta }{\delta \phi_0^*} 
\Biggr ) \textit{Z}
\nonumber\\&&
=
 \int d^Dx \, \Biggr (\frac{1}{\Lambda_D}
\frac{\delta \widehat\G}{\delta A^*_{a\mu}} 
\frac{\delta \widehat\G}{\delta A_a^\mu}
+\frac{1}{\Lambda_D}
\frac{\delta \widehat\G}{\delta \phi_a^*} 
\frac{\delta \widehat\G}{\delta \phi_a}
+ \frac{1}{\Lambda_D}
\frac{\delta \widehat\G}{\delta c_a^*}\frac{\delta \widehat\G}{\delta c_a}
\nonumber \\
&&
+ B_a \frac{\delta \G}{\delta \bar c_a} 
 + \Theta_{a\mu} \frac{\delta \widehat\G}{\delta V_{a\mu}}
      - K_0 \frac{\delta \widehat\G}{\delta \phi_0^*} 
\Biggr )\cdot \textit{Z}
\label{pert.7}
\end{eqnarray}
Thus the counterterms in perturbation theory must obey the 
following equation
\begin{eqnarray}&&\!\!\!
{\cal S}_0(\widehat\G^{(n)})+\frac{1}{\Lambda_D}
\sum_{j=1}^{n-1}\Biggr (
\frac{\delta \widehat\G^{(j)}}{\delta A^*_{a\mu}} 
\frac{\delta \widehat\G^{(n-j)}}{\delta A_a^\mu}
+
\frac{\delta \widehat\G^{(j)}}{\delta \phi_a^*} 
\frac{\delta \widehat\G^{(n-j)}}{\delta \phi_a}
+ 
\frac{\delta \widehat\G^{(j)}}{\delta c_a^*}
\frac{\delta \widehat\G^{(n-j)}}{\delta c_a}
\Biggr )=0.
\nonumber \\
\label{pert.8}
\end{eqnarray}
As in eq. (\ref{pert.4}) the nonlinear part of eq.
(\ref{pert.8}) fixes the violation at $n$ loops
of eq. (\ref{brst.13}) and therefore the implementability
of the pure pole subtraction strategy.
\subsection{Subtraction Procedure and ST Identity}
After the subtraction has been performed at $n-1$
order, the $n$-th order correction to the vertex functional
obeys the equation
\begin{eqnarray}
&&
\int d^Dx \,\Biggl[\frac{1}{\Lambda_D} \Big (
\frac{\delta \G^{(0)}}{\delta A^*_{a\mu}} \frac{\delta }{\delta A_a^\mu}
+
\frac{\delta \G^{(0)}}{\delta A_a^\mu}
\frac{\delta }{\delta A^*_{a\mu}} 
+
\frac{\delta \G^{(0)}}{\delta \phi_a^*} \frac{\delta }{\delta \phi_a}
+
\frac{\delta \G^{(0)}}{\delta \phi_a}
\frac{\delta }{\delta \phi_a^*} 
\nonumber \\
&& %~~~~~~~~ 
+
\frac{\delta \G^{(0)}}{\delta c_a^*}\frac{\delta }{\delta c_a}
+ 
\frac{\delta \G^{(0)}}{\delta c_a} \frac{\delta }{\delta c_a^*}\Big )
%\nonumber \\
%&& ~~~~~~~~
+ B_a \frac{\delta }{\delta \bar c_a}  + \Theta_{a\mu} \frac{\delta }{\delta V_{a\mu}}
      - K_0 \frac{\delta }{\delta \phi_0^*} 
 \Biggl]\G^{(n)}
\nonumber \\
&&  +\frac{1}{\Lambda_D}
\int d^Dx \, \, \sum_{j=1}^{n-1} \Big ( 
\frac{\delta \G^{(j)}}{\delta A^*_{a\mu}} \frac{\delta \G^{(n-j)}}{\delta A_a^\mu}
+
\frac{\delta \G^{(j)}}{\delta \phi_a^*} \frac{\delta \G^{(n-j)}}{\delta \phi_a}
+
\frac{\delta \G^{(j)}}{\delta c_a} \frac{\delta \G^{(n-j)}}{\delta c_a^*}
\Big )\nonumber \\
&& =\frac{1}{\Lambda_D}
\sum_{j=1}^{n-1}\Biggr (
\frac{\delta \widehat\G^{(j)}}{\delta A^*_{a\mu}} 
\frac{\delta \widehat\G^{(n-j)}}{\delta A_a^\mu}
+
\frac{\delta \widehat\G^{(j)}}{\delta \phi_a^*} 
\frac{\delta \widehat\G^{(n-j)}}{\delta \phi_a}
+ 
\frac{\delta \widehat\G^{(j)}}{\delta c_a^*}
\frac{\delta \widehat\G^{(n-j)}}{\delta c_a}
\Biggr ).
\label{pert.9}
\end{eqnarray}
In fact only after the introduction of the counterterm $
\widehat\G^{(n)}$ the ST identity is expected to be valid
by  eq. (\ref{pert.7}). The missing counterterm can be replaced
by the nonlinear part exhibited in eq. (\ref{pert.8}), thus yielding the
breaking term in the R.H.S. of eq. (\ref{pert.9}).
\par 
A closer look at the L.H.S. of eq. (\ref{pert.9}) shows that the operator
acting on  $\G^{(n)}$ does not contain $\Lambda_D$ 
and that the last nonlinear
terms involving $\G^{(j)} ~(j<n)$ have no pole parts by assumption.
We now divide both terms by  $\Lambda_D$ in order to normalize
the vertex functional in the L.H.S. according to eq. (\ref{pert.5.2})
and to remove any $D$-dependence in the R.H.S. apart from the pole in $D-4$,
in agreement with the normalization of the counterterms in
eq. (\ref{pert.3}). The subtraction of the poles from
$\Lambda_D^{-1}\G^{(n)}$ leaves invariant in form the L.H.S. of the
equation and the breaking terms are removed. 
I.e. one recovers the ST identity for the subtracted amplitudes at order $n$.

%}
%%%%%%%%%%%%%%%%%%%%%%%%%%%%%%%%%%%%%%%%%%%%%%%%%%%%%%
%
\section{Weak Power-Counting II}
%%%%%%%%%%%%%%%%%%%%%%%%%%%%%%%%%%%%%%%%%%%%%%%%%%%%%%
\label{sec:wpc2}

By making use of the functional equation associated to 
the local gauge transformations 
one can indeed establish a weak power-counting theorem.
The number of independent ancestor amplitudes can be fixed
by taking into account the functional identities which are
fulfilled by the vertex functional $\G$. 
As we have already discussed, the ST identity (\ref{brst.13})
is not enough to induce a hierarchy. 
%\par\noindent
The Landau gauge equation (\ref{b.eq})
shows that the dependence on $B$ only enters at tree level
(since the R.H.S. of this equation is purely classical).

The ghost equation (\ref{gh.eq})
fixes the dependence on $\bar c_a$. Therefore the field $\bar c_a$
can be neglected in the hierarchy procedure.

The functional equation (\ref{bkgwi})
will in turn fix the dependence on the $\phi$'s.
The ancestor amplitudes can correspondingly be identified with
those involving the ancestor variables, i.e. all the fields and sources
except the $\vec\phi$-fields.

The weak power-counting theorem can be stated as follows.
The number of independent superficially divergent amplitudes is
finite at each order in the loop expansion. These
amplitudes involve only the ancestor fields and sources.
In particular,
given a 1-PI $n$-loop graph ${\cal G}$ with $N_A$ external
$A$-legs, $N_c$ external $c$-legs,
$N_V$ external $V$-legs, $N_\Theta$ external $\Theta$-legs,
$N_{\phi_0^*}$ external $\phi_0^*$-legs, $N_{K_0}$ external $K_0$-legs,
$N_{\phi_a^*}$ external $\phi_a^*$-legs, $N_{A^*}$ external
$A^*$-legs and $N_{c^*}$ external $c^*$-legs, the superficial
degree of divergence of ${\cal G}$ is bounded by
\begin{eqnarray}
&& d({\cal G}) \leq (D-2)n +2 - N_A - N_c - N_V - N_{\phi_a^*} \nonumber \\
&& ~~~~~~~~ - 2 (N_\Theta + N_{A^*} + N_{\phi_0^*} + N_{c^*} + N_{K_0} ) \, . 
%\nonumber \\
\label{wk.9}
\end{eqnarray}
Moreover this property is stable under minimal subtraction
in dimensional regularization.
A detailed proof of this result is given in Appendix~\ref{app:C}.
>From eq.(\ref{wk.9}) we see that at each order in the loop expansion
there is only a finite number of divergent ancestor amplitudes.
\par\noindent
The above result relies on the assumptions discussed in Section
\ref{sec:unique}.
\par

Our subtraction scheme is consistent and predictive.
It is consistent since the defining equations are stable under
the subtraction procedure and it is predictive because the physical
parameters are those of the zero-loop vertex functional plus
the scale of the radiative corrections (denoted in the present paper
by $\Lambda$). 
Uniqueness of the tree-level vertex functional
(see Sect.~\ref{sec:unique}),
as dictated by the symmetries and the weak power-counting, 
forbids additional terms.
Physical unitarity in the Landau gauge has been proved
under quite general assumptions \cite{Ferrari:2004pd} 
and it is based on the ST identity (\ref{brst.13}).
By the weak power-counting the number of counterterms
is finite at each order in the loop expansion (see
eq. (\ref{pert.3})).
The $n$-th loop counterterms contain ancestor
monomials with dimension bounded by eq.(\ref{wk.9}).

\par
We finally notice that from eq. (\ref{wk.9}) one can associate a {\sl
  "dimension''} (distinct from the canonical dimension) which serves
to establish the degree of divergence of a graph. This allows to
establish a grading in the local solutions of the homogeneous
equations for the counterterms (eqs. (\ref{one.1})). This technique
will be used in Section \ref{sec:one} for the construction of a basis
for the counterterms in the one-loop approximation.

\section{Uniqueness of the Tree-Level Vertex Functional}
\label{sec:unique}

We are now in a position to prove the uniqueness of the tree-level
vertex functional in eq.(\ref{brst.4.1}).
The dependence on the antifields is fixed by the boundary conditions
in eq.(\ref{tl.cond}).
The dependence on $B_a$ and on the antighost field
$\bar c_a$ is determined by eqs.(\ref{b.eq})  and
(\ref{gh.eq}) respectively.
The local $SU(2)_L$ symmetry is implemented through
eq.(\ref{bkgwi}). Then the ST identity in eq.(\ref{brst.13}) 
fixes the dependence on $V_{a\mu}$ and $K_0$,
as well as on the ghosts $c_a$. 
However, by requiring global $SU(2)_R$ invariance
there is still the freedom to add any global $SU(2)_R$-invariant
constructed out of the bleached variable $a_{a\mu}$.
This residual freedom  is indeed limited by the 
weak power-counting theorem. For that purpose we first  notice
that only invariants
up to dimension $4$ in the $A_{a\mu}$ variables are allowed 
by the UV behavior of the $A_{a\mu}$-propagator.
Such an argument is shared also by power counting renormalizable
theories.
This limits the possible interactions terms to  the set of invariants 
in eq.(\ref{stck.2}).
Then the central idea of the
argument is that only $(G_{a\mu\nu}[a])^2$ is independent of the
fields $\vec\phi$ and $\phi_0$, in a way already shown in eq. (\ref{stck.1}).
If any dependence on the fields $\vec\phi$ and $\phi_0$
in the dimension four $a_\mu$ monomials in eq. (\ref{stck.2}) remains, 
then we get infinitely
many divergent graphs for the ancestor amplitudes 
already at one loop level (violation of
the weak power counting rule).

\begin{figure}
\begin{center}
\includegraphics[width=0.4\textwidth]{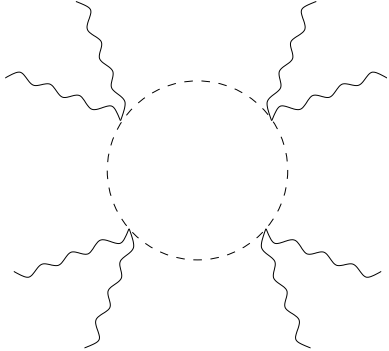}
\end{center}
\caption{A weak power-counting violating graph.}
\label{fig.1}
\end{figure}

First we notice that only the combination 
\begin{eqnarray}
\int d^Dx \, ( \partial_\mu a_{a\nu} \partial^\mu a^\nu_a - (\partial a_a)^2)
\label{uniq.1}
\end{eqnarray}
is allowed by the requirement of the absence of negative metric
modes in the $\phi_a$-sector. In fact if we expand 
$a_{a\mu}$ in powers of $\phi_a$ 
according to eq.(\ref{s2.5})
after setting the gauge field $A_{a\mu}$ to zero
we find at the lowest order
\begin{eqnarray}
&& 
\!\!\!\!\!\!\!\!\!\!\!\!\!\!\!\!\!\!\!\!\!\!\!
\int d^Dx \, \partial_\mu a_{a\nu} \partial^\mu a^\nu_a \sim
\int d^Dx \, (\partial a_a)^2 \sim
\frac{4}{v^2} \int d^Dx \, \phi_a \square^2 \phi_a \, .
\label{uniq.2}
\end{eqnarray} 
We now notice that each of the invariants in eq.(\ref{stck.2}),
with the exclusion of the mass term $\int d^4x~ a^2$,
contains vertices with two $A$'s, two $\phi$'s and two
derivatives. These vertices destroy the weak power-counting bound
in eq.(\ref{wk.9}) since they give rise already at one loop level
to divergent graphs involving an arbitrary number of
external $A$-legs (see Figure \ref{fig.1}) with superficial
degree of divergence $d({\cal G}) = 4$.

By requiring that these interaction vertices vanish
one finds that only the following combination is allowed
(up to an overall constant)
\begin{eqnarray}
%&& 
%\!\!\!\!\!\!\!\!\!\!\!\!\!\!
%\int d^Dx \, \Big ( \partial_\mu a_{a\nu} \partial^\mu a^\nu_a 
%- (\partial a_a)^2 +2 \epsilon_{abc} \partial_\mu a_{a\nu}
%a^\mu_b a^\nu_c + \frac{1}{2} (a^2)^2 - 
%\frac{1}{2} a_{a\mu} a_b^\mu a_{a\nu} a_b^\nu \Big ) \nonumber \\
%&& ~~~~~~~~~~~~~~~ =
\int d^Dx \, G_{a\mu\nu}[a] G^{\mu\nu}_a[a] \, .
\label{uniq.3}
\end{eqnarray}

This is a rather remarkable result.
The tree-level vertex functional in eq.(\ref{brst.4.1}) 
(which embodies the Yang-Mills action with
a St\"uckelberg mass term) is uniquely determined
by symmetry requirements and the 
weak power-counting property. In particular,
the symmetry content of the model allows
for the anomalous trilinear and quadrilinear couplings 
in eq.(\ref{stck.2}), but the latter are excluded
on the basis of weak power-counting criterion.

%%%%%%%%%%%%%%%%%%%%%%%%%%%%%%%%%%%%%%%%%%%%%%%%%%%%%%
%%%%%%%%%%%%%%%%%%%%%%%%%%%%%%%%%%%%%%%%%%%%%%%%%%%%%%
\section{One Loop}
%%%%%%%%%%%%%%%%%%%%%%%%%%%%%%%%%%%%%%%%%%%%%%%%%%%%%%
\label{sec:one}
%%%%%%%%%%%%%%%%%%%%%%%%%%%%%%%%%%%%%%%%%%%%%%%%%%%%%%
As it is well known there is no consistent theory
of pure massive Yang-Mills in the framework of 
power-counting renormalizable field theory (i.e. physical
unitarity is violated).  The present formulation
aims to overcome the limitation
of power-counting renormalizability and yet to provide a
consistent physical theory. This can be tested already
at one loop. In particular one can verify that the
conditions for the validity of physical unitarity are
met and moreover that the divergences can be consistently organized
in counterterms which preserve the defining equations
(symmetric subtraction). Some one-loop calculations
will be published elsewhere. Here we provide a theoretical
analysis of the counterterms by means of the local
solutions of  the linearized equations (\ref{pert.2})
and (\ref{pert.8}), that 
(together with the Landau gauge equation (\ref{b.eq}))
at one loop take the form
\begin{eqnarray}&&
{\cal W}_0(\widehat\G^{(1)}) =0
\nonumber\\&&
{\cal S}_0(\widehat\G^{(1)}) =0
\nonumber\\&&
\frac{\delta \widehat\G^{(1)}}{\delta B_a}=0.
\label{one.1}
\end{eqnarray}
We want to provide a basis for the local solutions
of eqs. (\ref{one.1}). The FP ghost number has to be zero
and moreover the dimensions of the monomial must match
those of the pole part of the Feynman amplitudes.
The analysis of this problem is made easy by the fact that
\begin{eqnarray}
\biggl[{\cal W}_0,
{\cal S}_0\biggr]=0.
\label{one.2}
\end{eqnarray}
This can be proved by the following line of steps:
i) the commutator is either zero or of first order in the
functional derivatives thus eq. (\ref{one.2}) needs to be
checked only on the fields and sources; ii) on fields eq. (\ref{one.2})
reduces to eq. (\ref{brst0.1}); iii) in order to test eq. (\ref{one.2})
on the sources we use the identities
\begin{eqnarray}
&&
{\cal W}_0(\G^{(0)})=-\frac{1}{2}\int d^Dx\alpha^L_a
\biggl(K_0\phi_a-\phi_0\frac{\delta\G^{(0)}}{\delta\phi_a}\biggr)
\nonumber\\&&
{\cal S}_0(\G^{(0)})= -\int d^Dx\biggl(B_a \frac{\delta}{\delta\bar c_a}
+\Theta_{a\mu}\frac{\delta}{\delta V_{a\mu}}
-K_0\frac{\delta}{\delta\phi^*_a}
\biggr)\G^{(0)}.
\label{one.3}
\end{eqnarray}
The equation (\ref{one.2}) is used in order to construct the
solutions of eqs. (\ref{one.1}). The solutions of
\begin{eqnarray}
{\cal W}_0({\cal M})=0
\label{one.4}
\end{eqnarray}
are constructed by using ``bleached'' fields. Then eq. (\ref{one.2})
says that the ST transform of ${\cal M}$ also satisfies eq. (\ref{one.4})
\begin{eqnarray}
{\cal W}_0\bigl( {\cal S}_0({\cal M})\bigr)
= 
{\cal S}_0\bigl( {\cal W}_0({\cal M})\bigr)=0.
\label{one.5}
\end{eqnarray}
In Appendix \ref{app:D} we explicitly realize eq. (\ref{one.5})
by showing that the ST transform by ${\cal S}_0$ of bleached
fields and sources remains bleached. 
The results are used here to  perform the ST transforms
on monomials that yields solutions of eqs. (\ref{one.1}) of
dimensions equal or less than four. The output of this
calculation provides a basis for the counterterms at the one loop
level. The invariants can be divided into two classes, one
depends only on the bleached gauge field $a_\mu$ and the rest
is given by all possible ${\cal S}_0$-exact local functional
of dimension less or equal four. The first class is given by

\begin{eqnarray}
&& {\cal I}_1 = \int d^Dx \, Tr ~ \partial_\mu a_\nu \partial^\mu a^\nu \, ,
\nonumber \\
&& {\cal I}_2 = \int d^Dx \, Tr ~ (\partial a)^2 \, , \nonumber \\
&& {\cal I}_3 = i\int d^Dx \, Tr ( \partial_\mu a_\nu [ a^\mu, a^\nu ] ) \, ,
\nonumber \\
&& {\cal I}_4 = \int d^Dx \, Tr  (a^2) ~ Tr (a^2)  \, , 
\nonumber \\
&& {\cal I}_5 = \int d^Dx \, Tr ( a_\mu a_\nu ) Tr  (a^\mu a^\nu ) \, ,
\nonumber \\
&& {\cal I}_6 = \int d^Dx \, Tr (a^2) \,  .
\label{ol.inv.1}
\end{eqnarray}
By explicit calculation one shows that
\begin{eqnarray}&&
\!\!\!\!\!\!\!\!\!\!\!\!\!\!
 {\cal I}_1 
-{\cal I}_2
-3 {\cal I}_3
+2{\cal I}_4
-2{\cal I}_5 
-M^2 {\cal I}_6
=-  \int d^Dx \, Tr \Big [
a^\mu \Big (  
 D^\rho G_{\rho \mu}[a]  + M^2 a_\mu \Big ) 
 \Big ] \,
\nonumber\\&&
=-\frac{g^2}{\Lambda^{D-4}}{\cal S}_0 Tr \int d^Dx \,
\widetilde{\widehat{A}^*}_\mu a^\mu 
\label{ident}
\end{eqnarray}
i.e. the invariants ${\cal I}_1 - {\cal I}_6$ are linearly
independent, but  ${\cal S}_0$-dependent (cohomologically dependent).
Moreover it should be reminded that ${\cal I}_1 - {\cal I}_6$
can be linearly  combined to reproduce $\int d^Dx \, (G_{a\mu\nu})^2$, 
which is independent from $\vec\phi$. Thus one of the invariants out
of ${\cal I}_1 - {\cal I}_6$ can be discarded in favor of the squared field 
strength. This has some advantages if one projects to amplitudes
involving the Goldstone $\vec\phi$-field. 
%}
\par
The second class contains the ${\cal S}_0$-exact local functionals
\begin{eqnarray}
&& 
{\cal I}_7 = {\cal S}_0 \int d^Dx \,  Tr(
\widetilde{{\widehat A}^*}_\mu v^\mu) \nonumber \\ 
&& ~~~ = \frac{\Lambda^{D-4}}{g^2}  \int d^Dx \, Tr \Big [
v^\mu \Big (  
 D^\rho G_{\rho \mu}[a]  + M^2 a_\mu \Big ) 
 \Big ] %\,  , \nonumber \\
%&& ~~~~~ 
- \int d^Dx \, Tr (\widetilde{{\widehat A}^*}_\mu \widetilde \Theta^\mu) 
\nonumber \\
&& ~~~~~~~~~~~~~ + \int d^Dx \, Tr \widetilde{{\widehat A}^*}_\mu
(D^\mu[v] 
\tilde c) 
\,  ,
\nonumber \\
&& 
{\cal I}_{8} = {\cal S}_0 \Biggl[\int d^Dx \, Tr ( \widetilde \Omega^*(x))
{\cal S}_0\int d^Dy \, Tr ( \widetilde \Omega^*(y))
\Biggr]
\nonumber \\
&&
~~~ =\int d^Dx \,\Biggl[\biggl(Tr ( \widetilde K)\biggr)^2
-\biggl(Tr ( \widetilde c~\widetilde \Omega^* )\biggr)^2
+2i Tr ( \widetilde K)Tr ( \widetilde c~\widetilde \Omega^* )
\Biggr]
\nonumber \\
&&
{\cal I}_{9} = {\cal S}_0\int d^Dx \,  Tr ( \widetilde \Omega^*)
Tr ( a^2 )
\nonumber \\
&& ~~~ =
- i \int d^Dx \, Tr (\widetilde c ~ \widetilde \Omega^*) Tr ( a^2)
- \int d^Dx \, Tr (\widetilde K) Tr( a^2)\, , 
\nonumber \\
&& {\cal I}_{10} =  {\cal S}_0\int d^Dx \, Tr ( \widetilde c^*
\widetilde c )
\nonumber \\
&& ~~~ = \int d^Dx \, \Big (  Tr ((D^\mu[a] \widetilde{{\widehat A}^*}_\mu) \widetilde c) - \frac{i}{4}
Tr ((\widetilde \Omega^*)^\dagger  \tilde c)
+ \frac{i}{2} Tr (\widetilde c^* \{ \widetilde c, \widetilde c \} )
\Big )  \, ,
 \nonumber \\
&& 
{\cal I}_{11} = {\cal S}_0 \int d^Dx \, Tr( \widetilde \Omega^*) =
- i \int d^Dx \, Tr (\widetilde c ~ \widetilde \Omega^*)
- \int d^Dx \, Tr( \widetilde K)\, . \nonumber \\
\label{ol.inv.2}
\end{eqnarray}
The last invariant ${\cal I}_{11}$, although of lower dimensions, 
has been included for a possible use in  gauges different from
Landau's. 

At the one-loop level the weak power-counting criterion
fixes the upper bound for the dimensions of the local invariants.
On the basis of this argument we have omitted invariants like:
\begin{eqnarray}&&
{\cal S}_0 \int d^Dx \, Tr ( \widetilde \Omega^*  \widetilde K )
=\int d^Dx Tr \biggl([- i \widetilde{c} ~ \widetilde \Omega^* 
- \widetilde K ]\widetilde K 
+i \widetilde \Omega^*~ \tilde c\widetilde K\biggr)
 \nonumber \\
&& 
= -\int d^Dx Tr \biggl( i \widetilde{c} ~ \{\widetilde \Omega^*, \widetilde K\}
+\widetilde K^2 \biggr),
\label{ol.inv.3}
\end{eqnarray}
since it has terms of dimension 5 according to  the counting 
of eq. (\ref{wk.9}).
%%%%%%%%%%%%%%%%%%%%%%%%%%%%%%%%%%%%%%%%%%%%%%
%%%%%%%%%%%%%%%%%%%%%%%%%%%%%%%%%%%%%%%%%%%%%%

\section{Conclusions}
\label{sec:concl}
%%%%%%%%%%%%%%%%%%%%%%%%%%%%%%%%%%%%%%%%%%%%%%
A consistent theory of massive Yang-Mills can
be formulated in spite of the fact that the
starting set of Feynman rules corresponds to a
power-counting nonrenormalizable theory.
Consistency is based on  the existence of
a subtraction scheme for the divergences which
does not alter the set of defining equations.
Physical unitarity, locality of the counterterms,
finite number of subtractions at each order
of the loop expansion (more correctly:
expansion in $\hbar$ ) and finite number of 
physical parameters are essential properties
of the procedure of subtraction. The symmetry
of the model is the gauge group SU(2)-left (local) $\otimes$
SU(2)-right (global). Moreover BRST invariance
is enforced in order to guarantee physical unitarity.
The managing of the divergences is based on
techniques already tested in the nonlinear sigma model:
hierarchy, weak power-counting and dimensional
subtraction on properly normalized 1-PI amplitudes.
The spontaneous breakdown of the global axial symmetry is
via a vacuum expectation value which has no physical
significance. The global vector symmetry remains unitarily
implemented.

%%%%%%%%%%%%%%%%%%%%%%%%%%%%%%%%%%%%%%%%%%%%%%%
%%%%%%%%%%%%%%%%%%%%%%%%%%%%%%%%%%%%%%%%%%%%%%%%
%%%%%%%%%%%%%%%%%%%%%%%%%%%%%%%%%%%%%%%%%%%%%%%
\appendix

\section{Feynman Rules}
\label{app:A}

In order to fix the Feynman rules we find it convenient to 
use the tree-level effective action
(\ref{pert.0.0}) instead of the original form in eq. (\ref{brst.4.2}).
By this choice  both the local functional equation (\ref{bkgwi}) and
the ST identity (\ref{brst.3}) acquire an explicit dependence on
$\Lambda_D=\Lambda^{D-4} /g^2$ (as discussed in Section \ref{sec:pert}).
\par
The advantage resides in the fact that with the rescaled
effective action (\ref{pert.0.0}) the dependence of the
1-PI amplitudes on $\Lambda_D$ can be easily traced:
any $n_L$-loop amplitude contains $\Lambda_D^{1-n_L}$ as a factor 
(see eq.(\ref{pert.5.3p})). Then one can 
discard any dependence from $\Lambda_D$ in the intermediate 
steps and recover it at the
end of the calculations. In particular when one evaluates the
counterterms, the prescription (\ref{pert.5.2}) requires that
at any loop order the amplitudes must be normalized 
by the prefactor $\Lambda^{4-D}$, before the
subtraction  of the poles (see eq. (\ref{pert.5.5})). 
On the other side, if physical
matrix elements are required, the normalization of the asymptotic
states has to be taken into account. Thus at  the tree level
approximation one gets for physical S-matrix elements
\begin{eqnarray}
S_{A_1 \dots A_{N_A}} = g^{N_A} W^C_{A_1 \dots A_{N_A}} 
\label{prop.1.1}
\end{eqnarray}
where $W^C_{A_1 \dots A_{N_A}}$ denotes
the connected amputated Green function with 
physical polarizations inserted on the gauge boson  legs
$A_1, \dots, A_{N_A}$. 
\par
The quadratic part in the quantized fields of $\G^{(0)}$ 
(where $\Lambda_D$ has been discarded) is
\begin{eqnarray}&&
\!\!\!\!\!\!\!\!\!
\int d^Dx \, \Biggl ( - \frac{1}{4} (\partial_\mu A_{a\nu} 
- \partial_\nu A_{a\mu})^2
+ \frac{M^2}{2} ( A_{a\mu} - \frac{2}{v} 
\partial_\mu \phi_a )^2 
+ B_a \partial A_a - \bar c_a \square c_a \Biggr )  .
\nonumber\\&&
\label{prop.1}
\end{eqnarray}
It is straightforward to get the propagators
\begin{eqnarray}
&&
\!\!\!\!\!\!\!\!\!\!\!\!\!\!\!\!\!\!
 \Delta_{A_{a\mu}A_{b\nu}} = \frac{-i }{p^2 -  M^2} \Big ( g_{\mu\nu} 
- \frac{p_\mu p_\nu}{p^2} \Big ) \delta_{ab} \, , 
~~~~
\Delta_{\phi_a \phi_b} 
= \frac{i}{4}\frac{ v^2}{ M^2} 
\frac{1}{p^2} \delta_{ab} \, , 
\nonumber \\&& 
\!\!\!\!\!\!\!\!\!\!\!\!\!\!\!\!\!\!
\Delta_{B_a A_{a\mu}} = \frac{p_\mu}{p^2} \delta_{ab} \, , ~~~~
\Delta_{B_a \phi_b} = - i \frac{v}{2 p^2} \, , ~~~~ \Delta_{c_a \bar c_b} 
= \frac{i}{p^2} \delta_{ab},
\nonumber \\&& 
\!\!\!\!\!\!\!\!\!\!\!\!\!\!\!\!\!\!
\Delta_{B_a B_b} =0,  ~~~~ \Delta_{A_{a\mu}\phi_b} =0\, .
\label{prop.2}
\end{eqnarray}

\section{Absence of the Hierarchy Based on Slavnov-Taylor Identity}\label{app:B}

At one loop order the ST identity in eq.(\ref{st.1}) reads
\begin{eqnarray}
&& 
\!\!\!\!\!\!\!\!\!\!\!\!\!\!\!\!\!\!\!\!\!\!\!\!\!\!
{\cal S}_0 (\G^{(1)}) =  \int d^Dx \, \Big ( 
\frac{\delta \G^{(0)}}{\delta A_{a\mu}^*} 
\frac{\delta \G^{(1)}}{\delta A_a^\mu} +
\frac{\delta \G^{(0)}}{\delta A_{a\mu}}
\frac{\delta \G^{(1)}}{\delta A_{a\mu}^*} +
\frac{\delta \G^{(0)}}{\delta \phi_a^*} 
\frac{\delta \G^{(1)}}{\delta \phi_a} +
\frac{\delta \G^{(0)}}{\delta \phi_a} 
\frac{\delta \G^{(1)}}{\delta \phi_a^*}
\nonumber \\
&& ~~~~~~~~~~~ + \frac{\delta \G^{(0)}}{\delta c_a^*} 
     \frac{\delta \G^{(1)}}{\delta c_a}
   + \frac{\delta \G^{(0)}}{\delta c_a}
     \frac{\delta \G^{(1)}}{\delta c_a^*}
   + B_a \frac{\delta \G^{(1)}}{\delta \bar c_a} \Big ) = 0 \, .
\label{appB:1}
\end{eqnarray}

In order to show that eq.(\ref{appB:1}) does not uniquely
fix the dependence on the $\phi$'s once
the amplitudes involving all the remaining variables
are known 
(absence of the hierarchy), 
 we will construct two different solutions 
 ${\cal I},{\cal I}'$ of eq.(\ref{appB:1})
which coincide at $\phi_a=0$.

For that purpose we notice that 
\begin{eqnarray}
{\cal I} & = & {\cal S}_0 ( \int d^Dx \, (A^*_{a\mu} + \partial_\mu \bar c_a) A^\mu_a)
\nonumber \\
& = & \int d^Dx \, \Big ( A_{a\mu} \frac{\delta S}{\delta A_{a\mu}}
- (A^*_{a\mu} + \partial_\mu \bar c_a) \partial^\mu c_a \Big ) 
\label{appB:2}
\end{eqnarray}
is ${\cal S}_0$-invariant due to the nilpotency of ${\cal S}_0$.
Nilpotency holds
as a consequence of the tree-level ST identity in eq.(\ref{brst.4}).
\par\noindent

At $\phi=0$ ${\cal I}$ reduces to
\begin{eqnarray}
{\cal I}_{\phi=0} & = & \int \, \Biggl[   \frac{\Lambda^{D-4}}{g^2}
\biggl ( - \partial_\mu A_{\nu a} \partial^\mu A^\nu_a 
       + (\partial A_a)^2 
\nonumber \\
& & ~~~~
- 3\epsilon_{abc} \partial_\mu A_{\nu a}A^\mu_b A^\nu_c 
- (A^2)^2 + A_{a\mu} A_b^\mu A_{a\nu} A_b^\nu  
%\nonumber \\
%& & ~~~~~ 
+ M^2 A_{\mu a}^2 \biggr)
\nonumber \\
& & ~~~~  
-   ( A^*_{\mu a} + \partial_\mu \bar c_a) \partial^\mu c_a \Biggr ]  .
\label{ctex.2}
\end{eqnarray}
We now set
\begin{eqnarray}
{\cal S}_0 \phi_a = \Omega_{ab} c_b
\equiv (\frac{1}{2} \phi_0 \delta_{ab} + \frac{1}{2} \epsilon_{acb} \phi_c )
c_b \, .
\label{trsf.1}
\end{eqnarray}
The matrix $\Omega_{ab}$ is invertible due to the nonlinear
constraint in eq.(\ref{s2.2}). 
Let us now replace in the first two lines of eq.(\ref{ctex.2})
$\partial_\mu$ with the covariant derivative w.r.t. $F_{a\mu}$. 
We substitute $A_{a\mu}$ with the combination $I_{a\mu}= A_{a\mu} - F_{a\mu}$.
Moreover we make separately ${\cal S}_0$-invariant the last line
of eq.(\ref{ctex.2}) as follows:
\begin{eqnarray}
{\cal I}' & = & \int \, \Biggl[  \frac{\Lambda^{D-4}}{g^2}
\biggl( - (D[F]_\mu I_\nu)_a (D[F]^\mu I^\nu)_a 
       + (D[F]I)_a^2 
\nonumber \\
& & ~~~~
- 3\epsilon_{abc} (D_\mu[F] I_\nu)_a I^\mu_b I^\nu_c 
%\nonumber \\
%&& ~~~~
- (I^2)^2 + I_{a\mu} I_b^\mu I_{a\nu} I_b^\nu  
%\nonumber \\
%& & ~~~~~ 
+ M^2 I^2 \biggl)
\nonumber \\
& & ~~~~  
+ {\cal S}_0\biggl[  ( A^*_{a\mu} + \partial_\mu \bar c_a) 
  \partial^\mu (\Omega^{-1}_{ap} \phi_p )\biggr] \Biggr] \, .
\label{ctex.4}
\end{eqnarray}
By construction ${\cal I}'$ is also ${\cal S}_0$-invariant.
Moreover at $\phi=0$ ${\cal I}$ and ${\cal I}'$ coincide, 
as can be easily checked
by noticing that
\begin{eqnarray}
&& 
\!\!\!\!\!\!\!\!\!\!
{\cal S}_0 ( ( A^*_{a\mu} + \partial_\mu \bar c_a) 
  \partial^\mu (\Omega^{-1}_{ap} \phi_p )) = 
{\cal S}_0  ( A^*_{a\mu}+ \partial_\mu \bar c_a ) \partial^\mu (
\Omega^{-1}_{ap} \phi_p ) \nonumber \\
& &  ~~~~~~~~~~  
- (A^*_{a\mu}+ \partial_\mu \bar c_a)
\partial^\mu \Big ( {\cal S}_0 ( \Omega^{-1}_{ap}) \phi_p \Big )
%\nonumber \\
%&&  ~~~~ 
-  (A^*_{a\mu}+ \partial_\mu \bar c_a)
\partial^\mu c_a \, .
\label{ctex.5}
\end{eqnarray}

However ${\cal I}$ and ${\cal I}'$ differ in their $\phi$-dependent terms.
Let us consider for instance 
the sector $A^* c \phi$. By using integration
by parts a basis of monomials involving
just one derivative 
is given by 
$\epsilon_{abc} \partial A^*_a  c_b \phi_c$, 
$\epsilon_{abc} A^*_{a\mu}  \partial^\mu c_b \phi_c$.
We project on the latter monomial.
The only term contributing to this monomial in ${\cal I}'$
is
\begin{eqnarray}
\frac{2}{v} A^*_{a \mu} \epsilon_{abc} \partial^\mu c_b \phi_c \, .
\label{ctex.12}
\end{eqnarray}
On the other hand, there is no similar contribution in
${\cal I}$ (since in ${\cal I}$ $A^*_{a\mu}$ does not couple to the $\phi$'s).

This means that we have found two different
 ${\cal S}_0$-invariants
with the same ancestor amplitudes.
This gives an explicit counterexample
showing that the ST identity is not sufficient
in order to implement the  hierarchy principle.

\section{Proof of the Weak-Power Counting Formula}
\label{app:C}

In this appendix we prove the power-counting formula in 
eq.(\ref{wk.9}).

Let ${\cal G}$ 
be an arbitrary $n$-loop 1-PI ancestor graph with $I$ internal lines, $V$ vertices and a given set $\{ N_A, N_c, N_V, N_\Omega, N_{\phi_0^*},
N_{K_0}, N_{\phi_a^*}, N_{A^*}, N_{c^*} \}$ of external legs.

By eq.(\ref{prop.2}) all propagators but those involving the field $B$
behave as $p^{-2}$ as $p$ goes to infinity, while
those involving $B$ behave as $p^{-1}$.
Let us denote by
${\hat I}$ the number of internal lines associated
with propagators which do not involve $B$, and by $I_B$ the number of 
internal lines with propagators involving $B$. 
One has 
\begin{eqnarray}
I= {\hat I} + I_B \, .
\label{wk.new}
\end{eqnarray}

According to the Feynman rules generated by
the tree-level vertex functional in eq.(\ref{brst.4.1})
the superficial degree of divergence of ${\cal G}$ is 
\begin{eqnarray}
&& \!\!\!\!\!\!\!\!\!\!\!\!\!\!\!\!\!\!\!\!\!\!
d({\cal G}) =  n D - 2 {\hat I} - I_B + V_{AAA} + \sum_k V_{A \phi^k} +
                     2 \sum_k V_{\phi^k} + V_{\bar c c A}
                                         + V_{\bar c c V} \, .
\label{wk.2}
\end{eqnarray}
In the above equation we have denoted by $V_{AAA}$ the number of 
vertices in ${\cal G}$ with three $A$-fields, 
with $V_{A\phi^k}$ the number of vertices with one $A$
and $k$ $\phi$'s and so on.
By using eq.(\ref{wk.new}) we can rewrite eq.(\ref{wk.2}) as
\begin{eqnarray}
\!\!\!\!\!\!\!\!\!\!\!\!\!\!\!\!
d({\cal G}) & = & n D - 2 I + I_B + V_{AAA} + \sum_k V_{A \phi^k} +
                     2 \sum_k V_{\phi^k} + V_{\bar c c A}
                                         + V_{\bar c c V} \, .
\label{wk.2.bis}
\end{eqnarray}
Moreover, since $B$ only enters into the trilinear
vertex $\G^{(0)}_{B_a V_{b\mu} A_{c\nu}}$, the number
of $BVA$ vertices must coincide with the number of propagators
involving $B$:
\begin{eqnarray} 
I_B = V_{BVA} \, .
\label{wk.2.ter}
\end{eqnarray}

The total number of vertices $V$ is given by
\begin{eqnarray}
V & = & V_{AAA} + V_{AAAA} + \sum_k V_{A \phi^k} + \sum_k V_{\phi^k} \nonumber \\
  &   & +V_{BVA} + V_{\bar c c A} + V_{\bar c c V}
        +V_{\bar c c V A}
\nonumber \\
  &   & + V_{\bar c A \Omega} +  V_{\phi_0^* \phi c}
        + \sum_k V_{\phi_a^* \phi^k c} \nonumber \\
  &   & + V_{A^* A c} + V_{c^* c c} 
        + \sum_k V_{K_0 \phi^k} \, .
\label{wk.3}
\end{eqnarray}

Euler's formula yields
\begin{eqnarray}
I = n + V -1 \, .
\label{wk.4}
\end{eqnarray}
By using eq.(\ref{wk.2.ter}), (\ref{wk.3}) and eq.(\ref{wk.4}) 
into eq.(\ref{wk.2}) one
gets
\begin{eqnarray}
d({\cal G}) & = & (D-2)n +2 + I_B \nonumber \\
     &   & - V_{AAA} - \sum_k V_{A\phi^k} -V_{\bar c c  A}
           - V_{\bar c c V} \nonumber \\
     &   & - 2 \Big [ V_{AAAA} + V_{BVA} + V_{\bar c c V A}
                     +V_{\bar c A \Omega} \nonumber \\
     &   & ~~ + V_{\phi_0^* \phi c}
                   + \sum_k V_{\phi_a^* \phi^k c} 
                   + V_{A^* A c} + V_{c^* c c } 
                   + \sum_k V_{K_0 \phi^k} \Big ] 
\nonumber \\
     & = & (D-2)n +2  \nonumber \\
     &   & - V_{AAA} - \sum_k V_{A\phi^k} -V_{\bar c c  A}
           - V_{\bar c c V} \nonumber \\
     &   & - V_{BVA} - 2 \Big [ V_{AAAA} + V_{\bar c c V A}
                     +V_{\bar c A \Omega} \nonumber \\
     &   & ~~ + V_{\phi_0^* \phi c}
                   + \sum_k V_{\phi_a^* \phi^k c} 
                   + V_{A^* A c} + V_{c^* c c } 
                   + \sum_k V_{K_0 \phi^k} \Big ] \, .
\label{wk.5}
\end{eqnarray}

Clearly one has
\begin{eqnarray}
&& V_{\bar c A \Omega} = N_{\Omega} \, , ~~~~ 
V_{\phi_0^* \phi c} = N_{\phi_0^*} \, , 
\nonumber \\
&& V_{A^* A c} = N_{A^*} \, , ~~~~
   V_{c^* c c} = N_{c^*} \, , \nonumber \\
&& \sum_k V_{\phi_a^* \phi^k c} = N_{\phi_a^*} \, , ~~~~
\sum_k V_{K_0 \phi^k} = N_{K_0} \, , \nonumber \\
&& V_{\bar c c V} +  V_{BVA} +  V_{\bar c c V A} = 
 N_V \, .
\label{wk.6}
\end{eqnarray}
Moreover
\begin{eqnarray}
&& V_{AAA} + \sum_k V_{A \phi^k} + 2 V_{AAAA} +  V_{\bar c c A}+  V_{\bar c c V A} + \sum_k V_{\phi_a^* \phi^k c} \nonumber \\
&& ~~~~ \geq N_A + N_c \, .
\label{wk.8}
\end{eqnarray}
In fact the quadrilinear vertex $V_{AAAA}$ can give one 
or two external $A$ lines.

By using eqs.(\ref{wk.6}) and (\ref{wk.8}) into
eq.(\ref{wk.5}) we obtain 
in a straightforward way the following bound:
\begin{eqnarray}
&& d({\cal G}) \leq (D-2)n +2 - N_A - N_c - N_V - N_{\phi_a^*} \nonumber \\
&& ~~~~~~~~ - 2 (N_\Omega + N_{A^*} + N_{\phi_0^*} + N_{c^*} + N_{K_0} ) \, .
\label{wk.9.bis}
\end{eqnarray}
This establishes the validity of the weak power-counting formula.
%%%%%%%%%%%%%%%%%%%%%%%%%%%%%%%%%%%
%%%%%%%%%%%%%%%%%%%%%%%%%%%%%%%%%%%%

\section{${\cal S}_0$-transforms of the Bleached Variables}
\label{app:D}

In this Appendix we derive the ${\cal S}_0$-transforms of
the bleached variables. For that purpose it is useful
to work in matrix notation. 

The ${\cal S}_0$-transform of $\Omega$ in eq.(\ref{s2.2})
is
\begin{eqnarray}
{\cal S}_0 \Omega = i c \Omega \, , 
\label{app.D.0}
\end{eqnarray}
where
\begin{eqnarray}
c = c_a \frac{\tau_a}{2} \, .
\label{app.D.0.1}
\end{eqnarray} 
Moreover
\begin{eqnarray}
{\cal S}_0 c = \frac{i}{2} \{ c, c \} \, .
\label{app.D.0.2}
\end{eqnarray}
It follows by direct computation that the bleached partner of $c$
\begin{eqnarray}
\widetilde c = \Omega^\dagger c \Omega 
\label{app.D.0.3}
\end{eqnarray}
transforms as follows under ${\cal S}_0$:
\begin{eqnarray}
{\cal S}_0 \widetilde c = - \frac{i}{2} \{ \widetilde{c}, \widetilde{c} \} \, .
\label{app.D.0.4}
\end{eqnarray}

$a_\mu$ in eq.(\ref{s2.5}) is ${\cal S}_0$-invariant.
On the other hand the ${\cal S}_0$-transform of
\begin{eqnarray}
v_\mu = v_{a\mu} \frac{\tau_a}{2} = \Omega^\dagger (V_{\mu} - F_{ \mu} ) \Omega 
\label{app.D.1}
\end{eqnarray}
yields
\begin{eqnarray}
{\cal S}_0 v_\mu = \Omega^\dagger ( \Theta_\mu 
- D_\mu[V] c )
\Omega 
=
\widetilde \Theta_\mu - D_\mu[v] \tilde c \, ,
\label{app.D.2}
\end{eqnarray}
and
\begin{eqnarray}
{\cal S}_0 \widetilde \Theta_\mu = -i \{ \widetilde c , \widetilde \Theta_\mu \} \, .
\label{app.D.3}
\end{eqnarray}

We now move to the study of the antifield-dependent
sector.

For that purpose we first evaluate 
the ${\cal S}_0$-variation of
\begin{eqnarray}
\widehat A^*_\mu = \widehat A^*_{a\mu} \frac{\tau_a}{2}
\label{app.D.5} 
\end{eqnarray}
and get according to eq.(\ref{gh.eq.3.l}):
\begin{eqnarray}
\!\!\!\!\!\!\!\!\!
{\cal S}_0 \widehat A^*_\mu & = & \frac{\delta S}{\delta A_{a\mu}}
\frac{\tau_a}{2} - %\Lambda^{D-4} 
\epsilon_{abc} c_b \widehat A^*_{c\mu} \frac{\tau_a}{2} \nonumber \\
 & = & \frac{\Lambda^{D-4}}{g^2} \Big [ 
 D^\rho G_{a \rho \mu} \frac{\tau_a}{2} + M^2 (A_{a\mu} - F_{a\mu})
 \frac{\tau_a}{2} \Big ] + i %\Lambda^{D-4} 
 \{ c, \widehat A^*_\mu
 \} \, .
 \label{app.D.6}
\end{eqnarray}
We need to express the R.H.S. of the above equation
in terms of bleached variables. The bleached counterpart of
$\widehat A^*_\mu$ is
\begin{eqnarray}
\widetilde{\widehat A^*}_\mu = \Omega^\dagger \widehat A^*_\mu \Omega \, .
\label{app.D.7}
\end{eqnarray}
The transition from $A_\mu$ to the bleached gauge field
$a_\mu$ is achieved by means of a $SU(2)_L$ gauge transformation of
parameters $\Omega$:
\begin{eqnarray}
A_\mu = \Omega a_\mu \Omega^\dagger + i \Omega \partial_\mu \Omega^\dagger \, .
\label{app.D.8}
\end{eqnarray}
Since the terms between square brackets in eq.(\ref{app.D.6})
transform in the adjoint representation under $SU(2)_L$
gauge transformations we get, by taking into account
eqs.(\ref{app.D.0.3}) and (\ref{app.D.7}) 
\begin{eqnarray}
\!\!\!\!\!\!\!\!\!
{\cal S}_0 \widehat A^*_\mu 
 & = & \frac{\Lambda^{D-4}}{g^2}  \Omega \Big [ 
 D^\rho G_{\rho \mu}[a]  + M^2 a_\mu 
 \Big ]  \Omega^\dagger + i %\Lambda^{D-4} 
 \Omega \{ \widetilde{c}, \widetilde{\widehat A^*}_\mu
 \} \Omega^\dagger \, .
 \label{app.D.9}
\end{eqnarray}
and finally
\begin{eqnarray}
\!\!\!\!\!\!\!\!\!
{\cal S}_0\widetilde{{\widehat A}^*}_\mu 
 & = & \frac{\Lambda^{D-4}}{g^2}  \Big [ 
 D^\rho G_{\rho \mu}[a]  + M^2 a_\mu 
 \Big ]   \, .
 \label{app.D.10}
\end{eqnarray}

The matrix $\Omega^*$ in eq.(\ref{brst.7.1})
has the following ${\cal S}_0$-transform
\begin{eqnarray}
{\cal S}_0 \Omega^* = - K_0 + i \frac{\delta \G^{(0)}}{\delta \phi_a} \tau_a = - (K_0 + i K_a \tau_a) \equiv  -K 
\label{app.D.11}
\end{eqnarray}
where we have introduced the notation
\begin{eqnarray}
K_a \equiv - \frac{\delta \G^{(0)}}{\delta \phi_a} \, .
\label{app.D.12}
\end{eqnarray}
Under local left multiplication $K$ transforms as $\Omega$ \cite{Ferrari:2005va}.
The bleached counterpart of $\Omega^*$ is 
\begin{eqnarray}
\widetilde{\Omega^*} = \Omega^\dagger \Omega^* \, .
\label{app.D.13}
\end{eqnarray}
Its ${\cal S}_0$-transform gives
\begin{eqnarray}&&
{\cal S}_0 \widetilde{\Omega^*} = - i \Omega^\dagger c \Omega^* 
- \Omega^\dagger K = - i \widetilde{c} ~ \widetilde \Omega^* - \widetilde K 
\nonumber \\
&&
{\cal S}_0 \widetilde K=-i \widetilde{c} ~\widetilde K\, .
\label{app.D.14}
\end{eqnarray}

Finally we consider the ${\cal S}_0$-variation of
\begin{eqnarray}
c^* = c_a^* \frac{\tau_a}{2} \, .
\label{app.D.15}
\end{eqnarray}
It is convenient to rewrite the couplings between the antifields
$(\phi_0^*,\phi_a^*)$ and the BRST variations
$(s \phi_0, s \phi_a)$ in eq.(\ref{brst.4.1}) 
in the following way
\begin{eqnarray}
&& \int d^Dx \, \Big ( \phi_0^* s\phi_0 + \phi_a^* s \phi_a \Big ) =
\int d^Dx \, \frac{1}{2} Tr [ (\Omega^*)^\dagger s \Omega ]
\nonumber \\
&& ~~~~~~~~~~~
=  \int d^Dx \, \frac{1}{2} Tr [ (\Omega^*)^\dagger i c_a \frac{\tau_a}{2} \Omega ] \, .
\label{app.D.16}
\end{eqnarray}
One finds
\begin{eqnarray}
{\cal S}_0 c^* & = & \frac{\delta \G^{(0)}}{\delta c_a} \frac{\tau_a}{2} =
D^\mu[A]\widetilde{\widehat A^*}_\mu -  \frac{i}{2} Tr [ (\Omega^*)^\dagger  \frac{\tau_a}{2} \Omega ]\frac{\tau_a}{2} - i [c^*,c] \nonumber \\
                            & = & \Omega (D^\mu[a]\widetilde{\widehat
                              A^*}_\mu) 
\Omega^\dagger -  \frac{i}{2} Tr [  (\widetilde \Omega^*)^\dagger \Omega^\dagger  \frac{\tau_a}{2} \Omega ] \frac{\tau_a}{2}   - i \Omega [\tilde c^*,\tilde c] \Omega^\dagger                        \, .
\label{app.D.17}
\end{eqnarray}
Then we consider the ${\cal S}_0$-variation of
\begin{eqnarray}
\widetilde c^* = \Omega^\dagger c^* \Omega
\label{app.D.18}
\end{eqnarray}
and we get
\begin{eqnarray}
{\cal S}_0 \widetilde c^* = (D^\mu[a]\widetilde{\widehat A^*}_\mu) 
-
 \frac{i}{2} Tr [  (\widetilde \Omega^*)^\dagger ~ \Omega^\dagger  \frac{\tau_a}{2} \Omega ]\Omega^\dagger  \frac{\tau_a}{2} \Omega \, .
\label{app.D.19}
\end{eqnarray}
Since the matrices ${\cal T}_a = \Omega^\dagger  \frac{\tau_a}{2} \Omega$
are unitarily equivalent to the Pauli matrices
the bleached matrix $(\widetilde \Omega^*)^\dagger$
can be decomposed as follows:
\begin{eqnarray}
(\widetilde \Omega^*)^\dagger = \frac{1}{2} Tr [(\widetilde \Omega^*)^\dagger] {\bf 1} + 2 ~ Tr [ (\widetilde \Omega^*)^\dagger
{\cal T}_a ]  {\cal T}_a 
\label{app.D.20}
\end{eqnarray}
and thus finally the R.H.S. of eq.(\ref{app.D.19}) can be rewritten
as
\begin{eqnarray}
{\cal S}_0 \widetilde c^* = (D^\mu[a]\widetilde{\widehat A^*}_\mu) 
- \frac{i}{4} (\widetilde \Omega^*)^\dagger + \frac{i}{8} Tr [(\widetilde \Omega^*)^\dagger] {\bf 1}  \, .
\label{app.D.19.bis}
\end{eqnarray}

The results of this Appendix are quite remarkable. 
The ${\cal S}_0$-transforms of bleached variables are bleached.
The ${\cal S}_0$-transform of the bleached antifield
$\widetilde{\widehat A^*}_\mu$ is the equation of motion of the original
St\"uckelberg action $S$ in the bleached gauge field $a_\mu$
(see eq.(\ref{app.D.10})).

%%%%%%%%%%%%%%%%%%%%%%%%%%%%%%%%%%%%%%%%%%%%%%

%%%%%%%%%%%%%%%%%%%%%%%%%%%%%%%%%%%%%%%%%%%%%%

\section{Dependence on $v$}
\label{app:modST}

In this Appendix we derive an extended ST identity
allowing to control the dependence of the 
Green functions on $v$ through cohomological methods.
For that purpose we allow $v$ to transform under ST
differential ${\cal S}_0$ according to
\begin{eqnarray}
{\cal S}_0 v = \theta \, , ~~~  {\cal S}_0\theta = 0 \, , 
\quad \theta^2 =0 .
\label{app.ST.2}
\end{eqnarray}
The ST identity (\ref{brst.13}) is then modified to
\begin{eqnarray}
&& \!\!\!\!\!\!\!\!\!\!
{\cal S}(\G) = \int d^Dx \, \Big (
\frac{\delta \G}{\delta A^*_{a\mu}} \frac{\delta \G}{\delta A_a^\mu}
+
\frac{\delta \G}{\delta \phi_a^*} \frac{\delta \G}{\delta \phi_a}
+ 
\frac{\delta \G}{\delta c_a^*}\frac{\delta \G}{\delta c_a}
+ B_a \frac{\delta \G}{\delta \bar c_a} \nonumber \\
&& ~~~~~~~~~~~~~~~~ + \Theta_{a\mu} \frac{\delta \G}{\delta V_{a\mu}}
      - K_0 \frac{\delta \G}{\delta \phi_0^*} 
\Big ) + \theta \frac{\partial \G}{\partial v} = 0 \, .
\label{app.ST.5}
\end{eqnarray}
The effective action at the tree level $\G^{(0)}$ ((\ref{brst.4.1}) and
(\ref{brst.4.2})) is a solution
of the above equation only after adding an extra term dependent
on $v$ and $\theta$
\begin{eqnarray}
\G^{(0)}_{ext} & = & 
 S + 
\frac{\Lambda^{D-4}}{g^2}
\int d^D x \, \Big ( B_a (D^\mu[V](A_\mu - V_\mu))_a
- \bar c_a (D^\mu[V] D_\mu[A] c)_a \Big ) 
\nonumber \\& & 
+ \frac{\Lambda^{D-4}}{g^2} \int d^Dx \, \Theta_{a}^\mu~(D_\mu[A]\bar{c})_a 
\nonumber \\& & 
+ \int d^Dx \, \Big ( A^*_{a\mu} sA^\mu_a + 
\phi_0^* s~\phi_0  + 
\phi_a^* s~\phi_a
+ c_a^* s c_a
+ K_0 \phi_0 
\nonumber \\
& & 
+ \phi_0^* \frac{\theta}{v} \phi_0 + 
 \phi_a^* \frac{\theta}{v} \phi_a \Big ) \, .
\label{app.ST.4}
\end{eqnarray}
Now we can discuss the dependence of the physical amplitudes
from the parameter $v$. For this purpose it is convenient 
to introduce the connected generating
functional $W$ (we use the same notations as in eq.(\ref{pert.1}))
\begin{eqnarray}
W = \G + \int d^Dx \, \Big ( L_{a\mu} A^\mu_a +
K_a \phi_a + J^B_a B_a + \eta_a \bar c_a + \bar \eta_a c_a \Big )
\, .
\label{app.ST.6.0}
\end{eqnarray}
The ST identity for $W$ reads
\begin{eqnarray}
&& \!\!\!\!\!\!\!\!\!\!
{\cal S}(W) = \int d^Dx \, \Big (
-L_{a\mu} \frac{\delta W}{\delta A^*_{a\mu}}
-K_a
\frac{\delta W}{\delta \phi_a^*} 
- \bar \eta_a
\frac{\delta W}{\delta c_a^*}
- \frac{\delta W}{\delta J^B_a} \eta_a \nonumber \\
&& ~~~~~~~~~~~~~~~~ + \Theta_{a\mu} \frac{\delta W}{\delta V_{a\mu}}
      - K_0 \frac{\delta W}{\delta \phi_0^*} 
\Big ) + \theta \frac{\partial W}{\partial v} = 0 \, .
\label{app.ST.6}
\end{eqnarray}
This equation can be used in order to study the dependence
of the Green functions on $v$. In particular let
$\beta_{i_1}(x) , \dots , \beta_{i_n}(x_n)$ 
denote a set of additional external sources coupled
to BRST-invariant local operators 
${\cal O}_{i_1}(x_1) , \dots , {\cal O}_{i_n}(x_n)$.
By differentiating eq.(\ref{app.ST.6}) w.r.t. $\theta$
and $\beta(x_{i_1}), \dots, \beta(x_{i_n})$ and by setting all sources 
(collectively denoted by $\zeta$) to zero one gets
\begin{eqnarray}
\left . \frac{\partial}{\partial v} 
\frac{\delta^n W}{\delta \beta_{i_1}(x_1) \dots \delta \beta_{i_n}(x_n)} 
\right |_{\zeta=0} = 0 \, ,
\label{app.ST.7}
\end{eqnarray}
i.e. the Green functions of the operators ${\cal O}_i(x_i)$
are $v$-independent.
Moreover by differentiating eq.(\ref{app.ST.6}) w.r.t.
$\theta$  and $K_0$ we get
\begin{eqnarray}
\left . \frac{\partial}{\partial v} \frac{\delta W}{\delta K_0(x)} 
\right |_{\zeta=0} = 
\left . 
\frac{\partial}{\partial \theta} 
\frac{\delta W}{\delta \phi_0^*(x)} \right |_{\zeta=0} \, .
\label{app.ST.8}
\end{eqnarray}
This equation is a consequence of the fact that $\phi_0^*$
and $-K_0$ form a ${\cal S}_0$-doublet (see eq.(\ref{brst.12})).
We remark that a device technically  similar to the one 
adopted here (pairing of $v,\theta$ into a ${\cal S}_0$-doublet)
has been used in the context of gauge theories
in order to discuss the dependence on the gauge parameter.
However we stress an important difference: in the present case 
the dependence on $v$ is not confined to the
BRST-exact sector of the tree-level vertex functional,
since it also enters through the combination
$\frac{\phi_a}{v}$ in the St\"uckelberg mass term and in the
term $K_0\phi_0$ of (\ref{app.ST.4}).
Therefore $v$ cannot be identified {\em tout court}
with a  kind of gauge parameter.

\end{document}